\documentclass[runningheads]{llncs}

\usepackage[T1]{fontenc}
\usepackage{comment}
\usepackage{amsmath, amssymb, wasysym, stmaryrd, relsize, mathtools}
\usepackage{graphicx}
\usepackage[colorinlistoftodos]{todonotes}
\usepackage[colorlinks=true, allcolors=blue]{hyperref}
\usepackage{scalerel}
\usepackage{algorithm}
\usepackage[noend]{algpseudocode}
\usepackage{tikz}
\usetikzlibrary{arrows, arrows.meta, automata, shapes, positioning, decorations.pathreplacing}
\usepackage{float, subcaption}
\usepackage{xifthen}
\usepackage{textcomp, mathcomp}
\usepackage{tabularx}
\usepackage{fancyvrb}
\usepackage{centernot}

\makeatletter
\providecommand*{\twoheadrightarrowfill@}{%
  \arrowfill@\relbar\relbar\twoheadrightarrow
}
\providecommand*{\xtwoheadrightarrow}[2][]{%
  \ext@arrow 0579\twoheadrightarrowfill@{#1}{#2}%
}
\makeatother

\makeatletter
\providecommand*{\triangleheadrightarrowfill@}{%
  \arrowfill@\relbar\relbar\rightarrowtriangle
}
\providecommand*{\xtriangleheadrightarrow}[2][]{%
  \ext@arrow 0579\triangleheadrightarrowfill@{#1}{#2}%
}
\makeatother

\newcommand{\trans}[2][]{\xrightarrow{#2}\mathrel{\vphantom{\to}_{#1}}}
\newcommand{\strans}[2][]{\xrightarrow{#2}\mathrel{\vphantom{\to}_{#1}^*}}

\newcommand{\ptrans}[2][]{\xrightarrow{#2}\mathrel{\vphantom{\to}_{#1}^\prime}}

\newcommand{\Dtrans}[2][]{\xtriangleheadrightarrow{#2}\mathrel{\vphantom{\to}_{#1}}}

\newcommand{\concat}{\mathbin{+\mkern-10mu+}}

\newcommand{\cgtc}{\succ}

\newcommand{\corr}{\rhd}
\newcommand{\unre}{\blacktriangleright}
\newcommand{\chan}[2]{[#2]_{#1}}

\newcommand{\IL}{\ensuremath{I\!L}}
\newcommand{\Conf}{\ensuremath{Con\!f}}
\newcommand{\SC}[1]{\ensuremath{?SC_{C'}^C\!(#1)}}
\renewcommand{\AC}[2]{\ensuremath{!AC_{C'}^C\!(#1, #2)}}

\DeclareMathOperator{\biginterl}{\scalerel*{||}{\sum}}

\tikzset{class/.style={draw, rectangle, minimum height=0.5cm, minimum width=0.5cm}}
\tikzset{state/.style={draw, circle, minimum height=0.5cm, minimum width=0.5cm, align=center, inner sep = 0.5mm}}
\tikzset{component/.style={draw, rectangle, minimum height=0.75cm, minimum width=0.75cm}}
\tikzset{func/.style={->,>=triangle 60, auto, thick}}
\tikzset{comp/.style={<-,>=diamond, auto, thick}}
\tikzset{edge/.style={-, thick}}
\tikzset{inher/.style={->,>={Latex[angle'=90, open, scale=1.5]}, thick}}
\tikzset{lts/.style={->,>={Latex[angle'=60]}, auto, thick}}
\tikzset{fill right half/.style={path picture={\fill[#1] (path picture bounding box.north) rectangle (path picture bounding box.south east);}}}
\tikzset{queuer/.style={draw=none, rectangle, minimum height=0.5cm, minimum width=0.75cm, append after command={
            [very thick]
            (\tikzlastnode.north west) edge[-] (\tikzlastnode.north east)
            (\tikzlastnode.north west) edge[-] (\tikzlastnode.south west)
            (\tikzlastnode.south west) edge[-] (\tikzlastnode.south east)
        }
    }
}
\tikzset{queuel/.style={draw=none, rectangle, minimum height=0.5cm, minimum width=0.75cm, append after command={
            [very thick]
            (\tikzlastnode.north west) edge[-] (\tikzlastnode.north east)
            (\tikzlastnode.north east) edge[-] (\tikzlastnode.south east)
            (\tikzlastnode.south west) edge[-] (\tikzlastnode.south east)
        }
    }
}

\newcolumntype{R}{>{\raggedleft\arraybackslash}X}

\newcommand{\hvoucher}{\includegraphics[scale=0.11]{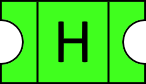}}
\newcommand{\uvoucher}{\includegraphics[scale=0.11]{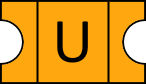}}
\newcommand{\apple}{\includegraphics[scale=0.02]{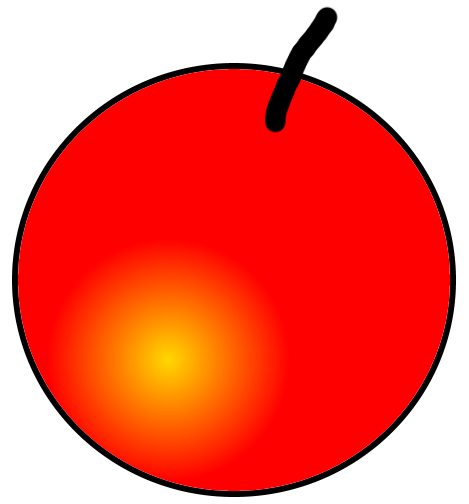}}
\newcommand{\banana}{\includegraphics[scale=0.032]{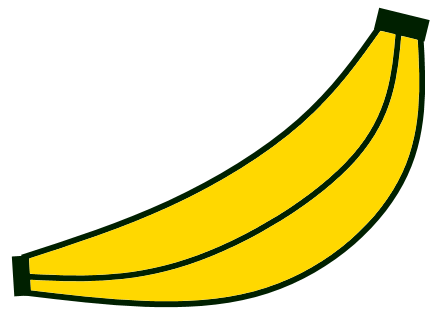}}
\newcommand{\chocolate}{\includegraphics[scale=0.055]{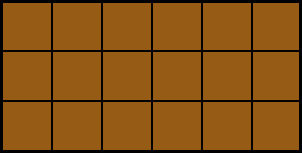}}
\newcommand{\donut}{\includegraphics[scale=0.035]{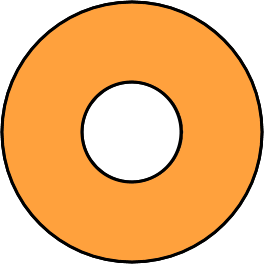}}

\usepackage[normalem]{ulem}

\title{On the Preservation of Properties when Changing Communication Models %
\thanks{This work was carried out as part of the VOICE-B project, which is funded by Canon Production Printing.}}

\author{Olav Bunte\inst{1} \and
Louis C.M. van Gool\inst{2} \and
Tim A.C. Willemse\inst{1}}
\authorrunning{O. Bunte et al.}

\institute{Eindhoven University of Technology, Eindhoven, Netherlands\\
\email{\{o.bunte, t.a.c.willemse\}@tue.nl} \and
Canon Production Printing, Venlo, Netherlands\\
\email{louis.vangool@cpp.canon}}

\begin{document}

\maketitle

\begin{abstract}
In a system of processes that communicate asynchronously by means of FIFO channels, there are many options in which these channels can be laid out.
In this paper, we compare channel layouts in how they affect the behaviour of the system using an ordering based on splitting and merging channels.
This order induces a simulation relation, from which the preservation of safety properties follows.
Also, we identify conditions under which the properties reachability, deadlock freedom and confluence are preserved when changing the channel layout.

\end{abstract}

\section{Introduction}

In asynchronous communication, sending and receiving a message are two separate actions, which makes it possible for messages to be received in a different order than they were sent.
What orderings are possible, depends on the asynchronous communication model(s) used within the system, for which many flavours are possible.
We consider communication models that are implemented by means of a layout of (unbounded) FIFO (First In First Out) channels, which defines how messages in transit are stored.
For instance, using a channel per message implements a fully asynchronous model, while having a single input channel per process enforces that messages that are sent to the same process are received in the same order in which they are sent.

While (re)designing or refactoring a software system of asynchronously communicating processes, it may be desirable to change (part of) the channel layout.
This can, for instance, be the case when design choices are still being explored, when the performance of the system needs to be improved, when the behaviour of the system has grown too complex due to the additions of new processes, or when the channel implementation is part of legacy software.
However, changing the channel layout may impact the behaviour of the system in unexpected ways, possibly violating desired properties.
In this paper, we investigate the extent of this impact.

We use the notion of a FIFO system \cite{DBLP:journals/lmcs/FinkelP20,DBLP:conf/concur/BolligFS20} to represent a software system of asynchronously communicating processes.
Firstly, we define an ordering on FIFO systems based on whether one can be created from the other by merging channels.
We then analyse the difference in the behaviour between related FIFO systems and show that it induces a simulation order, from which the preservation of safety properties follows.
Secondly, we analyse whether reachability, deadlock freedom and confluence are preserved when changing the channel layout.
Reachability is particularly relevant in practice, since changing the method of communication should typically not cause previously possible process behaviour to become impossible.
If deadlock freedom is preserved, it is ensured that changing the method of communication does not introduce undesired situations where all processes are stuck waiting for each other.
Confluence is related to the independence of actions, which is often expected between actions of different processes.
A violation of confluence between actions of different processes indicates a possible race condition, where the faster process determines how the system progresses.
We identify conditions under which these properties are guaranteed to be preserved when merging or splitting channels.

\paragraph{Related work.}
In \cite{DBLP:journals/scp/EngelsMR02}, seven distinct channel-based asynchronous communication models are related to each other in a hierarchy based on trace and MSC implementability.
The authors of \cite{DBLP:journals/fac/ChevrouHQ16} also consider the causal communication model \cite{DBLP:journals/cacm/Lamport78}, and show a similar hierarchy.
They prove this hierarchy correct in \cite{DBLP:conf/fm/ChevrouH0Q19} using automated proof techniques.
Compared to these works, we consider mixed (channel-based) communication models, which is more realistic for complex software systems.
For communication models that can be defined by FIFO systems, the hierarchies in these works relate these models the same way as our relation does.

Property preservation is closely related to the field of incremental model checking \cite{DBLP:conf/cav/SokolskyS94,DBLP:conf/birthday/HenzingerJMS03,DBLP:conf/tacas/WijsE13}, which is an efficient method for rechecking a property on a system that has undergone some changes.
Under some conditions, one can actually prove that a property will be preserved, as shown in multiple contexts \cite{DBLP:journals/fmsd/LoiseauxGSBB95,DBLP:conf/fmoods/Wehrheim00,DBLP:conf/memocode/HuangVG03,DBLP:journals/fac/DerrickS12,DBLP:journals/ieeejas/XiaL21}.
To our knowledge, no such work exists in the context of asynchronously communicating processes however.

\paragraph{Outline.}
We first introduce the necessary definitions to reason about FIFO systems in Section \ref{sec:preliminaries}.
We define an ordering between FIFO systems in Section \ref{sec:layouts} and show that it induces a simulation relation.
Then in Section \ref{sec:preservation} we identify conditions under which the aforementioned properties are preserved when changing the channel layout.
Lastly, we conclude in Section \ref{sec:conclusion}.
The proofs for all lemmas and theorems in Section \ref{sec:layouts} and \ref{sec:preservation} can be found in the Appendix.

\section{The FIFO system}
\label{sec:preliminaries}

Let $P$ be a set of processes that make up a software system and let $M$ be the set of messages that can be communicated between these processes.
We represent the behaviour of each process $p \in P$ with a Labelled Transition System (LTS) $B_p = \langle Q_p, q_p^0, L_p, \Dtrans[p]{}\rangle$ where $Q_p$ is its set of states, $q_p^0$ its initial state, $L_p \subseteq (\{?, !\} \times M) \cup \{\tau\}$ its set of actions and $\Dtrans[p]{} \;\subseteq Q_p \times L_p \times Q_p$ its transition relation.
An action $?m$ indicates the receiving of $m$, $!m$ the sending of $m$ and $\tau$ is an internal action.
We assume that processes do not share non-internal actions, that is $L_p \cap L_{p'} \subseteq \{\tau\}$ for all distinct $p, p' \in P$.
We write $q \Dtrans[p]{a} q'$ iff $(q, a, q') \in \;\Dtrans[p]{}$.
A FIFO system then describes how these processes communicate with each other via FIFO channels.

\begin{definition}
A \emph{FIFO system} is a tuple $\langle P, C, M \rangle$ where $C \subseteq \mathcal{P}(M)$ is a set of (FIFO) channels defined as a partition of $M$.
\end{definition}

Each channel in $C$ is defined as a set of messages, which represents the messages that this channel can hold.
Note that because $C$ partitions $M$, we assume that each message can only be sent to and received from exactly one channel.
For a message $m \in M$, we define $[m]_C$ as the channel in $C$ that $m$ belongs to, that is $m \in [m]_C$ and $[m]_C \in C$.
We write $m \simeq_C o$ iff $[m]_C = [o]_C$ for messages $m, o \in M$.

We define $M^*$ as the set of all finite sequences of messages, also known as \emph{words}.
We use $\epsilon$ as the empty word and concatenate two words with $\concat$.
Given a word $m \concat w$ for message $m \in M$ and word $w \in M^*$, we define its head as $hd(m \concat w) = m$ and its tail as $tl(m \concat w) = w$.

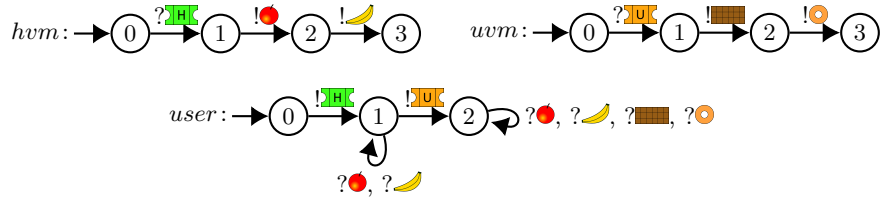
\begin{figure}[b]
\begin{subfigure}{0.5\textwidth}
\centering
\begin{tikzpicture}[lts]

\node		(name)	at	(-1.2, 0.04)	{$hvm\!:$};
\coordinate	(init)	at	(-0.75, 0);
\node[state]	(0)		at	(0, 0)		{$0$};
\node[state]	(1)		at	(1.2, 0)	{$1$};
\node[state]	(2)		at	(2.4, 0)	{$2$};
\node[state]	(3)		at	(3.6, 0)	{$3$};

\path
	(init)	edge								(0)
	(0)		edge	node	{$?\hvoucher{}$}	(1)
	(1)		edge	node	{$!\apple{}$}		(2)
	(2)		edge	node	{$!\banana{}$}		(3);

\end{tikzpicture}
\end{subfigure}\hfill%
\begin{subfigure}{0.5\textwidth}
\centering
\begin{tikzpicture}[lts]

\node		(name)	at	(-1.2, 0.01)	{$uvm\!:$};
\coordinate	(init)	at	(-0.75, 0);
\node[state]	(0)		at	(0, 0)		{$0$};
\node[state]	(1)		at	(1.2, 0)	{$1$};
\node[state]	(2)		at	(2.4, 0)	{$2$};
\node[state]	(3)		at	(3.6, 0)	{$3$};

\path
	(init)	edge						(0)
	(0)		edge	node	{$?\uvoucher{}$}	(1)
	(1)		edge	node	{$!\chocolate{}$}	(2)
	(2)		edge	node	{$!\donut{}$}		(3);

\end{tikzpicture}
\end{subfigure}\\[1em]
\begin{subfigure}{\textwidth}
\centering
\begin{tikzpicture}[lts]

\node		(name)	at	(-1.2, 0)	{$user\!:$};
\coordinate	(init)	at	(-0.75, 0);
\node[state]	(0)		at	(0, 0)		{$0$};
\node[state]	(1)		at	(1.2, 0)	{$1$};
\node[state]	(2)		at	(2.4, 0)	{$2$};

\path
	(init)	edge	(0)
	(0)		edge				node	{$!\hvoucher{}$}			(1)
	(1)		edge[loop below]	node	{$?\apple{}$, $?\banana{}$}	(1)
			edge				node	{$!\uvoucher{}$}			(2)
	(2)		edge[loop right]	node	{$?\apple{}$, $?\banana{}$, $?\chocolate{}$, $?\donut{}$}	(2);

\end{tikzpicture}
\end{subfigure}

\caption{Processes $hvm$, $uvm$ and $user$ for Example \ref{exa:vending1} and \ref{exa:vending2}.}
\label{fig:vendingproc}
\end{figure}

\begin{example}
\label{exa:vending1}
Imagine two vending machines, one for healthy snacks and one for unhealthy snacks, and some user who can interact with these vending machines.
After receiving a "healthy voucher" \hvoucher{}, the healthy vending machine can supply apples \apple{} and bananas \banana{}.
After receiving an "unhealthy voucher" \uvoucher{}, the unhealthy vending machine can supply chocolate \chocolate{} and donuts \donut{}.
The user decides to use \hvoucher{} before \uvoucher{} and can receive the snacks whenever they are ready.

Let $P_V = \{hvm, uvm, user\}$ be processes that represent the two vending machines and the user.
Their LTSs are visualised in Figure \ref{fig:vendingproc}.
The set of messages is $M_V = \{\hvoucher{}, \uvoucher{}, \apple{}, \banana{}, \chocolate{}, \donut{}\}$.
The realistic case where both vending machines have their own voucher slot and output slot is represented by the channel set $C_V = \{\{\hvoucher{}\}, \{\uvoucher{}\}, \{\apple{}, \banana{}\}, \{\chocolate{}, \donut{}\}\}$, resulting in the FIFO system $\langle P_V, C_V, M_V \rangle$.
Note that $\apple{} \simeq_{C_V} \banana{}$ and $\chocolate{} \simeq_{C_V} \donut{}$.
\end{example}

\paragraph{Semantics}
A FIFO system induces an LTS that represents the communication behaviour between all processes.
A state in this LTS consists of two parts: the states of the individual processes and the contents of the channels.
For a set of processes $P$, $\mathbf{P} = \{\kappa \in P \rightarrow \bigcup_{p \in P} Q_p\ |\ \forall_{p \in P} : \kappa(p) \in Q_p\}$ denotes the set of functions that map processes to their current states.
For a set of channels $C$, $\mathbf{C} = \{\zeta \in C \rightarrow M^*\ |\ \forall_{c \in C} : \zeta(c) \in c^*\}$ denotes the set of functions  that map channels to their contents.
In case of a set of channels $C'$, we write $\mathbf{C}'$.
Note that we assume unbounded channels.

\begin{definition}
\label{def:FIFO2LTS}
Let $F = \langle (Q_p, q_p^0, \Dtrans[p]{})_{p \in P}, C, M \rangle$ be a FIFO system.
The \emph{semantics} of $F$ is an LTS $B_F = \langle S, s_0, L, \trans{}\rangle$ where
\begin{itemize}
\item $S = \mathbf{P} \times \mathbf{C}$,
\item $s_0 = (\kappa_0, \zeta_\epsilon)$, where $\kappa_0(p) = q_p^0$ for all $p \in P$ and $\zeta_\epsilon(c) = \epsilon$ for all $c \in C$,
\item $L = (\{?, !\} \times M) \cup \{\tau\}$,
\item $\trans{}\; \subseteq S \times L \times S$ such that for all $(\kappa, \zeta) \in S$, $p \in P$, $q \in Q_p$ and $m \in M$, with $c = \chan{C}{m}$:
\begin{equation*}
\begin{split}
(\kappa, \zeta) \trans{\tau} (\kappa[p \mapsto q], \zeta) &\text{ iff } \kappa(p) \Dtrans[p]{\tau} q\\
(\kappa, \zeta) \trans{?m} (\kappa[p \mapsto q], \zeta[c \mapsto tl(\zeta(c)]) &\text{ iff } \kappa(p) \Dtrans[p]{?m} q \wedge hd(\zeta(c)) = m\\
(\kappa, \zeta) \trans{!m} (\kappa[p \mapsto q], \zeta[c \mapsto \zeta(c) \concat m]) &\text{ iff } \kappa(p) \Dtrans[p]{!m} q
\end{split}
\end{equation*}
\end{itemize}
\end{definition}

We write $s \trans{a} s'$ iff $(s, a, s') \in\; \trans{}$.
We lift the transition relation to one over sequences of actions $\strans{} \;\subseteq S \times L^* \times S$ in the usual way.
In the context of a FIFO system $F$, we refer to the semantics of the FIFO system as defined above as ``the LTS of $F$''.

Two LTSs can be compared by means of a simulation relation \cite{DBLP:journals/fmsd/LoiseauxGSBB95}.

\begin{definition}
Let $B = \langle S, s_0, L, \trans{}\rangle$ and $B' = \langle S', s_0', L, \ptrans{}\rangle$ be two LTSs.
We say that $B$ \emph{simulates} $B'$ iff there exists a \emph{simulation relation} $R \subseteq S' \times S$ such that $s_0'Rs_0$ and for all $s \in S$ and $s' \in S'$, if $s'Rs$ and $s' \ptrans{a} t'$ for some $t' \in S'$ and $a \in L$, then there must exist a $t \in S$ such that $s \trans{a} t$ and $t'Rt$.
\end{definition}

\section{Comparing channel layouts}
\label{sec:layouts}

The choice in channel layout affects the behaviour of a FIFO system.
The more channels there are, the more orderings there are in which messages can be received.
With this in mind, we order FIFO systems as follows:

\begin{definition}
\label{def:cgtc}
Let $F = \langle P, C, M \rangle$ and $F' = \langle P, C', M \rangle$ be two FIFO systems.
We define the relation $\cgtc$ on FIFO systems such that $F \cgtc F'$ iff $C \neq C'$ and $\forall_{c \in C} : \exists_{c' \in C'} : c \subseteq c'$ (that is, $C$ is a more refined partition of $M$ than $C'$ is).
\end{definition}

One can create $F'$ from $F$ by merging a number of channels (splitting channels in the opposite direction).
We first illustrate how this affects the behaviour of the system with an example.

\begin{example}
\label{exa:vending2}
Continuing from Example \ref{exa:vending1}, consider the FIFO systems $F_m = \langle P_V, \{\{\hvoucher{}\}, \{\uvoucher{}\}, \{\apple{}\}, \{\banana{}\}, \{\chocolate{}\}, \{\donut{}\}\}, M_V \rangle$ (one channel per message), $F_o = \langle P_V, \{\{\hvoucher{}, \uvoucher{}\}, \{\apple{}, \banana{}\}, \{\chocolate{}, \donut{}\}\}, M_V \rangle$ (one output channel per process) and $F_g = \langle P_V, \{M_V\}, M_V \rangle$ (one global channel).
Observe that $F_m \cgtc F_o \cgtc F_g$.

In $F_m$, the trace $!\hvoucher{}!\uvoucher{}?\uvoucher{}$ is possible, but in $F_o$ it is not.
This is because in $F_o$, both vouchers sent by $user$ are put in the same channel, so $hvm$ has to receive its voucher before $uvm$ can.
In $F_o$, the trace $!\hvoucher{}!\uvoucher{}?\hvoucher{}?\uvoucher{}!\apple{}!\chocolate{}?\chocolate$ is possible, but in $F_g$ it is not.
This is because in $F_g$, both vending machines send their snacks to the same channel, which fixes the order in which $user$ receives the snacks.
\end{example}

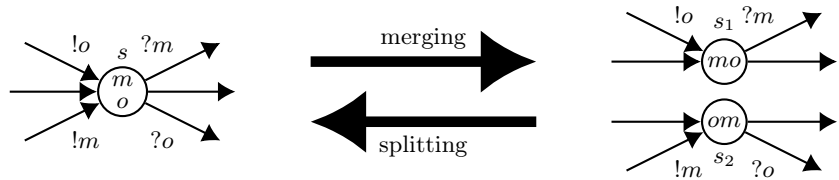
\begin{figure}[b]
\centering
\begin{tikzpicture}[lts]

\node[state]	(l)		at	(0, 0)		{$m$\\[-1mm]$o$};
\node			(s)		at	(0, 0.5)	{$s$};
\node[state]	(rt)	at	(8, 0.4)	{$mo$};
\node			(s1)	at	(8, 0.9)		{$s_1$};
\node[state]	(rb)	at	(8, -0.4)	{$om$};
\node			(s2)	at	(8, -0.9)		{$s_2$};

\coordinate	(lpin)		at	(-1.5, 0);
\coordinate	(lmin)		at	(-1.3, -0.65);
\coordinate	(loin)		at	(-1.3, 0.65);
\coordinate	(lpout)		at	(1.5, 0);
\coordinate	(lmout)		at	(1.3, 0.65);
\coordinate	(loout)		at	(1.3, -0.65);
\coordinate	(rtpin)		at	(6.5, 0.4);
\coordinate	(rtoin)		at	(6.7, 1.05);
\coordinate	(rtpout)	at	(9.5, 0.4);
\coordinate	(rtmout)	at	(9.3, 1.05);
\coordinate	(rbpin)		at	(6.5, -0.4);
\coordinate	(rbmin)		at	(6.7, -1.05);
\coordinate	(rbpout)	at	(9.5, -0.4);
\coordinate	(rboout)	at	(9.3, -1.05);
\coordinate	(mal)		at	(2.5, 0.4);
\coordinate	(mar)		at	(5.5, 0.4);
\coordinate	(sal)		at	(2.5, -0.4);
\coordinate	(sar)		at	(5.5, -0.4);

\path	(lpin)	edge						(l)
		(lmin)	edge[swap]	node	{$!m$}	(l)
		(loin)	edge		node	{$!o$}	(l)
		(l)		edge						(lpout)
		(l)		edge		node	{$?m$}	(lmout)
		(l)		edge[swap]	node	{$?o$}	(loout)
		(rtpin)	edge						(rt)
		(rtoin)	edge		node	{$!o$}	(rt)
		(rt)	edge						(rtpout)
		(rt)	edge		node	{$?m$}	(rtmout)
		(rbpin)	edge						(rb)
		(rbmin)	edge[swap]	node	{$!m$}	(rb)
		(rb)	edge						(rbpout)
		(rb)	edge[swap]	node	{$?o$}	(rboout)
		(mal)	edge[line width = 4pt]	node	{merging}		(mar)
		(sar)	edge[line width = 4pt]	node	{splitting}		(sal);

\end{tikzpicture}
\caption{A visualisation of the effect on the LTS of a FIFO system when merging or splitting channels.
Transitions without a label cover any transition that is not already represented by other incoming or outgoing transitions.}
\label{fig:mergesplit}
\end{figure}

In the remainder of this section, to avoid duplication in definitions, lemmas and theorems, we universally quantify over FIFO systems $F = \langle P, C, M \rangle$ and $F' = \langle P, C', M \rangle$ such that $F \cgtc F'$, with $B_F = \langle S, s_0, L, \trans{} \rangle$ and $B_{F'} = \langle S', s_0', L, \ptrans{} \rangle$.

The effect on the LTS when changing the channel layout is visualised in Figure \ref{fig:mergesplit}.
When the channels $\{m\}$ and $\{o\}$ are merged into one channel $\{m, o\}$, state $s$ results in states $s_1$ and $s_2$, one for every interleaving of the contents of the two channels in $s$.
Conversely, when channel $\{m, o\}$ is split into channels $\{m\}$ and $\{o\}$, the channel contents of states $s_1$ and $s_2$ are split as well, making them coincide, resulting in $s$.
We say that state $s$ \emph{generalises} states $s_1$ and $s_2$ and that states $s_1$ and $s_2$ \emph{specialise} state $s$.

To define this formally, we first define the interleavings of words.
Given a message $m \in M$ and a set of words $W$, let $m \concat W = \{m \concat w\ |\ w \in W\}$.
Then for a set of words $W$, we define the set of possible interleavings of these words $\biginterl W$ as $\biginterl W = \{\epsilon\}$ if $W = \{\epsilon\}$, else $\biginterl W = \bigcup_{m \concat w \in W} m \concat \biginterl ((W \setminus \{m \concat w\}) \cup \{w\})$.

\begin{example}
\label{exa:vendingil}
Continuing from Example \ref{exa:vending2}, let $W = \{\epsilon, \apple\banana, \chocolate\donut\}$.
Then $||W = \{\apple\banana\chocolate\donut, \apple\chocolate\banana\donut, \apple\chocolate\donut\banana, \chocolate\apple\banana\donut, \chocolate\apple\donut\banana, \chocolate\donut\apple\banana\}$.
\end{example}

With this, we can define generalisation/specialisation of states as follows:

\begin{definition}
\label{def:chancorr}
Let $\zeta \in \mathbf{C}$.
For channels $C'$ we define the set of functions $\mathbf{C}_{\zeta}'$, each representing possible interleavings of channel contents in $\zeta$, as:
\[
\mathbf{C}_{\zeta}' = \Big{\{}\zeta' \in \mathbf{C}'\ \Big{|}\ \forall_{c' \in C'} : \zeta'(c') \in \biginterl\{\zeta(c)\ |\ c \in C \wedge c \subseteq c'\}\Big{\}}
\]
\end{definition}

\begin{definition}
\label{def:statecorr}
Let $s = (\kappa, \zeta) \in S$ and $s' = (\kappa', \zeta') \in S'$.
We say that $s$ \emph{generalises} $s'$ and $s'$ \emph{specialises} $s$, written as $s \corr s'$, iff $\kappa = \kappa' \wedge \zeta' \in \mathbf{C}_{\zeta}'$.
\end{definition}

\begin{example}
Continuing from Example \ref{exa:vending2}, take the LTSs of FIFO systems $F_o$ and $F_g$.
Let $\kappa \in \mathbf{P}_V$ such that $\kappa(hvm) = 3$, $\kappa(uvm) = 3$ and $\kappa(user) = 2$ (the vending machines have supplied their snacks).
Assume that $user$ has not retrieved any snack from the channels yet.
In $F_o$ there only exists one state $s = (\kappa, \zeta)$ that represents this situation, namely where $\zeta(\{\hvoucher, \uvoucher\}) = \epsilon$, $\zeta(\{\apple, \banana\}) = \apple\banana$ and $\zeta(\{\chocolate, \donut\}) = \chocolate\donut$.
In $F_g$ there are 6 such states, because the two vending machines use the same channel for output (the only channel $M$), so their outputs get interleaved.
Let $S'_\kappa$ be the set of these 6 states.
Let $(\kappa, \zeta') \in S'_\kappa$, then the possible values for $\zeta'(M_V)$ are the interleavings mentioned in Example \ref{exa:vendingil}.
The states in $S'_\kappa$ specialise state $s$ since they are stricter in how the snacks are ordered in the channel(s).
Vice versa, state $s$ generalises the states in $S'_\kappa$.
\end{example}

For every state in the LTS of a FIFO system, specialising or generalising states exist in the LTS of $\cgtc$-related FIFO systems.

\begin{lemma}
\label{lem:statecorr}
$\forall_{s' \in S'} : \exists_{s \in S} : s \corr s'$ and $\forall_{s \in S} : \exists_{s' \in S'} : s \corr s'$
\end{lemma}

As shown in Figure \ref{fig:mergesplit}, after merging channels $\{m\}$ and $\{o\}$, the action $?m$ is only possible from $s_1$, since it has the interleaving where $m$ is at the head of the channel.
Action $!m$ can only result in $s_2$, since it has the interleaving where $m$ is at the back of the channel.
Similar arguments can be made for $?o$ and $!o$.
Any other transitions to and from $s$ are possible for both $s_1$ and $s_2$.
When splitting channel $\{m, o\}$ the opposite happens: the incoming and outgoing transitions for $s$ are all transitions to and from $s_1$ and $s_2$ combined.

We show which transitions are preserved in the LTS when changing the channel layout formally in the below four lemmas, for any $\kappa_1, \kappa_2 \in \mathbf{P}$, $\zeta \in \mathbf{C}$, $\zeta' \in \mathbf{C}_{\zeta}'$ and $m \in M$, with $c = \chan{C}{m}$ and $c' = \chan{C'}{m}$.
Firstly, internal actions are always possible from specialising or generalising states after merging or splitting channels.

\begin{lemma}
\label{lem:transcorrtau}
$(\kappa_1, \zeta) \trans{\tau} (\kappa_2, \zeta)$ iff $(\kappa_1, \zeta') \ptrans{\tau} (\kappa_2, \zeta')$.
\end{lemma}

Input actions remain possible from the generalising state after splitting channels.
When merging channels, such actions are not possible from specialising states that do not have the required message at the head, which may be the case when the channel of this message was merged, as was illustrated by Figure \ref{fig:mergesplit}.

\begin{lemma}
\label{lem:transcorrrecnomerge}
If $c = c'$, then $(\kappa_1, \zeta) \trans{?m} (\kappa_2, \zeta[c \mapsto tl(\zeta(c))])$ iff $(\kappa_1, \zeta') \ptrans{?m} (\kappa_2, \zeta'[c' \mapsto tl(\zeta'(c'))])$.
\end{lemma}

\begin{lemma}
\label{lem:transcorrrecmerge}
If $c \neq c'$, then $(\kappa_1, \zeta) \trans{?m} (\kappa_2, \zeta[c \mapsto tl(\zeta(c))]) \wedge hd(\zeta'(c')) = m$ iff $(\kappa_1, \zeta') \ptrans{?m} (\kappa_2, \zeta'[c' \mapsto tl(\zeta'(c'))])$.
\end{lemma}

Output actions are always possible from specialising or generalising states after merging or splitting channels.
Note however that sending a message to a merged channel increases the number of possible interleavings, so not all specialising target states are reached, as was illustrated by Figure \ref{fig:mergesplit}.

\begin{lemma}
\label{lem:transcorrsend}
$(\kappa_1, \zeta) \trans{!m} (\kappa_2, \zeta[c \mapsto \zeta(c) \concat m])$ iff $(\kappa_1, \zeta') \ptrans{!m} (\kappa_2, \zeta'[c' \mapsto \zeta'(c') \concat m])$.
\end{lemma}

Note that for each of the above four lemmas, the target state of the $\trans{}$-transition generalises the target state of the $\ptrans{}$-transition.
Since the structure of the transitions in these lemmas is the same as in Definition \ref{def:FIFO2LTS}, it follows that the above lemmas cover all transitions in $\trans{}$ and $\ptrans{}$.
In general, merging channels reduces the behaviour that a FIFO system allows.
This can be formalised with the simulation preorder.

\begin{lemma}
\label{lem:corrsim}
$\corr^{-1}$ is a simulation relation.
\end{lemma}

\begin{theorem}
\label{the:simulation}
$B_F$ simulates $B_{F'}$.
\end{theorem}

\section{Property preservation}
\label{sec:preservation}

In the previous section we have formally shown how the LTS of a FIFO system is affected when changing the channel layout.
In this section we investigate how properties of a system are affected by such changes.
Here the question is: if a property $\phi$ holds on the LTS of a FIFO system $F$, denoted by $B_F \models \phi$, under which conditions does it still hold after changing the channel layout?
For this, we define the following notions.

\begin{definition}
\label{def:preservation}
Let $F$ and $F'$ be two FIFO systems such that $F \cgtc F'$ and let $\phi$ be some property on FIFO systems.
We say that:
\begin{itemize}
\item $\phi$ is \emph{merge-preserved} iff $B_F \models \phi \Rightarrow B_{F'} \models \phi$.
\item $\phi$ is \emph{split-preserved} iff $B_F \models \phi \Leftarrow B_{F'} \models \phi$.
\end{itemize} 
\end{definition}

In \cite{DBLP:journals/fmsd/LoiseauxGSBB95} it has already been shown that simulation preserves safety properties, that is properties of the form "some bad thing is not reachable", so from Theorem \ref{the:simulation} we can derive the following:

\begin{theorem}
\label{the:safetymerge}
Safety properties are merge-preserved.
\end{theorem}

In the remainder of this section, we analyse the preservation of reachability, deadlock freedom and confluence.
To avoid duplication in definitions, lemmas and theorems, we again universally quantify over FIFO systems $F = \langle P, C, M \rangle$ and $F' = \langle P, C', M \rangle$ such that $F \cgtc F'$, with $B_F = \langle S, s_0, L, \trans{} \rangle$ and $B_{F'} = \langle S', s_0', L, \ptrans{} \rangle$.

\subsection{Reachability}

Reachability asks whether a state can be reached in the LTS of a FIFO system by a sequence of transitions, starting from the initial state.

\begin{definition}
\label{def:reachable}
Let $B = \langle S, s_0, L, \trans{}\rangle$ be an LTS and let $L' \subseteq L$.
A state $s \in S$ is \emph{$L'$-reachable} in $B$ iff there exists a sequence of actions $\alpha \in L^{\prime *}$ such that $s_0 \strans{\alpha} s$.
We define $Reach_{L'}(S)$ as the set of all $L'$-reachable states.
We omit $L'$ if $L' = L$.
We define $Reach(B)$ as the LTS $B$ restricted to only reachable states and the transitions between them.
\end{definition}

Preservation of reachability depends on whether a state's specialising or generalising states are still reachable after changing the channel layout.
When splitting channels this is the case, as it follows from Theorem \ref{the:simulation}.

\begin{lemma}
\label{lem:reachsplit}
Let $s \in S$ and $s' \in S'$.
Assume that $s \corr s'$.
Then $s \in Reach(S) \Leftarrow s' \in Reach(S')$.
\end{lemma}

When merging channels however, reachability is only guaranteed to be preserved when only transitions have been taken with actions that do not use merged channels.
Formally, we define the set of such actions as $\IL(F, F') = \{\tau\} \cup \{?m, !m\ |\ m \in M \wedge \chan{C}{m} = \chan{C'}{m}\}$.

\begin{lemma}
\label{lem:reachmerge}
Let $s \in S$ and $s' \in S'$.
Assume that $s \corr s'$.
Then $s \in \break Reach_{\IL(F, F')}(S) \Rightarrow s' \in Reach_{\IL(F, F')}(S')$.
\end{lemma}

We argue using Figure \ref{fig:mergesplit} why other actions violate merge-preservation of reachability.
The transition with action $!m$ can be done to $s$ and to $s_2$, but not to $s_1$, because it does not have $m$ at the end of its channel.
If $s$ would not have any other incoming transitions, $s_2$ is possibly unreachable.
The transition with action $?m$ can be done from $s$ to some state $t$ (not depicted in the figure) and from $s_1$, but not from $s_2$ since it does not have $m$ at the head of its channel.
Due to this, some states that specialise $t$ are possibly unreachable.

Lifting these results to the full system, we will only focus on the reachability of process states.
For a $\kappa \in \mathbf{P}$ and $L' \subseteq L$, we say that $\kappa$ is ($L'$-)reachable in $F$ iff there exists a $\zeta \in \mathbf{C}$ such that $(\kappa, \zeta)$ is ($L'$-)reachable in $B_F$.

\begin{theorem}
\label{the:reachsplit}
For all $\kappa \in \mathbf{P}$, reachability of $\kappa$ is split-preserved.
\end{theorem}

\begin{theorem}
\label{the:reachmerge}
For all $\kappa \in \mathbf{P}$, $\IL(F, F')$-reachability of $\kappa$ is merge-preserved.
\end{theorem}

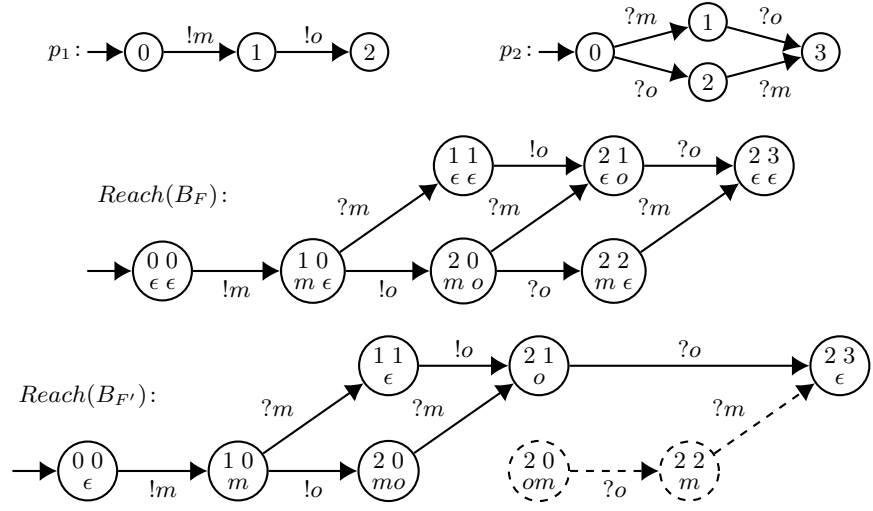
\begin{figure}[t!]
\begin{subfigure}[b]{\textwidth}
\centering
\begin{tikzpicture}[lts]

\node		(name)	at	(-1.05, -0.02)	{$p_1\!:$};
\coordinate	(init)	at	(-0.75, 0);
\node[state]	(0)	at	(0, 0)		{$0$};
\node[state]	(1)	at	(1.5, 0)	{$1$};
\node[state]	(2)	at	(3, 0)		{$2$};

\path
	(init)	edge					(0)
	(0)		edge	node	{$!m$}	(1)
	(1)		edge	node	{$!o$}	(2);

\node		(name)	at	(4.95, -0.02)	{$p_2\!:$};
\coordinate	(init)	at	(5.25, 0);
\node[state]	(0)	at	(6, 0)			{$0$};
\node[state]	(1)	at	(7.5, 0.4)		{$1$};
\node[state]	(2)	at	(7.5, -0.4)		{$2$};
\node[state]	(3)	at	(9, 0)			{$3$};

\path
	(init)	edge								(0)
	(0)		edge	node[pos=0.7]		{$?m$}	(1)
			edge	node[swap, pos=0.7]	{$?o$}	(2)
	(1)		edge	node[pos=0.3]		{$?o$}	(3)
	(2)		edge	node[swap, pos=0.3]	{$?m$}	(3);

\end{tikzpicture}
\end{subfigure}\\[1em]
\begin{subfigure}[b]{\textwidth}
\centering
\begin{tikzpicture}[lts]

\node		(name)	at	(0, 1)	{$Reach(B_F)\!:$};
\coordinate	(init)	at	(-1, 0);
\node[state]	(0)	at	(0, 0)		{$0 \; 0$\\[-1mm]$\epsilon \; \epsilon$};
\node[state]	(1)	at	(2, 0)		{$1 \; 0$\\[-1mm]$m \; \epsilon$};
\node[state]	(2)	at	(4, 0)		{$2 \; 0$\\[-1mm]$m \; o$};
\node[state]	(3)	at	(4, 1.4)	{$1 \; 1$\\[-1mm]$\epsilon \; \epsilon$};
\node[state]	(4)	at	(6, 0)		{$2 \; 2$\\[-1mm]$m \; \epsilon$};
\node[state]	(5)	at	(6, 1.4)	{$2 \; 1$\\[-1mm]$\epsilon \; o$};
\node[state]	(6)	at	(8, 1.4)	{$2 \; 3$\\[-1mm]$\epsilon \; \epsilon$};

\path
	(init)	edge							(0)
	(0)		edge	node[swap]		{$!m$}	(1)
	(1)		edge	node[swap]		{$!o$}	(2)
			edge	node[pos=0.4]	{$?m$}	(3)
	(2)		edge	node[swap]		{$?o$}	(4)
			edge	node[pos=0.4]	{$?m$}	(5)
	(3)		edge	node			{$!o$}	(5)
	(4)		edge	node[pos=0.4]	{$?m$}	(6)
	(5)		edge	node			{$?o$}	(6);

\end{tikzpicture}
\end{subfigure}\\[1em]
\begin{subfigure}[b]{\textwidth}
\centering
\begin{tikzpicture}[lts]

\node		(name)	at	(0, 1)	{$Reach(B_{F'})\!:$};
\coordinate	(init)				at	(-1, 0);
\node[state]			(0)		at	(0, 0)		{$0 \; 0$\\[-1mm]$\epsilon$};
\node[state]			(1)		at	(2, 0)		{$1 \; 0$\\[-1mm]$m$};
\node[state, dashed]	(2a)	at	(6, 0)		{$2 \; 0$\\[-1mm]$om$};
\node[state]			(2b)	at	(4, 0)		{$2 \; 0$\\[-1mm]$mo$};
\node[state]			(3)		at	(4, 1.4)	{$1 \; 1$\\[-1mm]$\epsilon$};
\node[state, dashed]	(4)		at	(8, 0)		{$2 \; 2$\\[-1mm]$m$};
\node[state]			(5)		at	(6, 1.4)	{$2 \; 1$\\[-1mm]$o$};
\node[state]			(6)		at	(10, 1.4)	{$2 \; 3$\\[-1mm]$\epsilon$};

\path
	(init)	edge									(0)
	(0)		edge			node[swap]		{$!m$}	(1)
	(1)		edge			node[swap]		{$!o$}	(2b)
			edge			node[pos=0.4]	{$?m$}	(3)
	(2a)	edge[dashed]	node[swap]		{$?o$}	(4)
	(2b)	edge			node[pos=0.4]	{$?m$}	(5)
	(3)		edge			node			{$!o$}	(5)
	(4)		edge[dashed]	node[pos=0.4]	{$?m$}	(6)
	(5)		edge			node			{$?o$}	(6);

\end{tikzpicture}
\end{subfigure}

\caption{Processes $p_1$ and $p_2$ and LTSs $Reach(B_F)$ and $Reach(B_{F'})$ for Example \ref{exa:reach}, with $F = \langle \{p_1, p_2\}, \{\{m\}, \{o\}\}, \{m, o\} \rangle$ and $F' = \langle \{p_1, p_2\}, \{\{m, o\}\}, \{m, o\} \rangle$.
The dashed states are unreachable states that specialise states in $Reach(B_F)$.}
\label{fig:reach}
\end{figure}

\begin{example}
\label{exa:reach}
See Figure \ref{fig:reach} for an example that shows that reachability of process states is not merge-preserved in general.
In $Reach(B_F)$, state $2$ of process $p_2$ is reachable, but in $Reach(B_{F'})$ it is not.
This is because in $Reach(B_{F'})$, the messages $m$ and $o$ can only be received by $p_2$ in the order in which they are sent by $p_1$.
Note that the actions $!m$, $!o$ and $?o$ that are necessary to reach state $2$ of $p_2$ in $Reach(B_F)$ are not in $\IL(F, F')$.
\end{example}

The preservation of reachability is not only interesting on its own, but also for property preservation in general, because one is typically only interested in the preservation of a property within reachable behaviour.
Thanks to Theorem \ref{the:reachsplit} and Lemma \ref{lem:statecorr}, we know that for merge-preservation of a property that needs to hold for all reachable states, it suffices to check whether if it holds for a state, then it also holds for its specialising states.
We cannot give such local arguments for split-preservation, because after splitting channels, states may be reachable that do not generalise any reachable state in the original LTS.
To be able to claim split-preservation of a property, we need to assume that all reachable states in the new system generalise some state in the original.
We call this assumption \emph{unaltered reachability} and represent it formally with $S \unre S'$, which is true iff for all $s \in Reach(S)$ there exists an $s' \in Reach(S')$ such that $s \corr s'$.

\subsection{Deadlock freedom}

A deadlock is a state in an LTS from which no action is possible.

\begin{definition}
Let $B = \langle S, s_0, L, \trans{}\rangle$ be an LTS.
A state $s \in S$ is a \emph{deadlock}, denoted as $\delta(s)$, iff there does not exist an $a \in L$ and $t \in S$ such that $s \trans{a} t$.
We say that $B$ is \emph{deadlock free} iff for all $s \in Reach(S)$ it holds that $\neg\delta(s)$.
\end{definition}

The preservation of deadlock freedom comes down to whether for every non-deadlock state, its generalising or specialising states are not deadlocks as well.
Whether this is the case can be easily derived from Figure \ref{fig:mergesplit}.
When splitting channels, the number of outgoing transitions cannot decrease, so $s$ cannot become a deadlock if $s_1$ or $s_2$ were not deadlocks already.

\begin{lemma}
\label{lem:deadlocksplit}
Let $s \in S$ and $s ' \in S'$.
Assume that $s \corr s'$.
Then $\neg\delta(s) \Leftarrow \neg\delta(s')$.
\end{lemma}

When merging channels however, the number of outgoing transitions with input actions may decrease in specialising states, which can cause some to become a deadlock.
There does always exist a specialising state that is not a deadlock, namely one where the interleaving of channel contents is such that the input action is still possible, but much more than this cannot be shown.
For instance, referring to Figure \ref{fig:mergesplit}, if only $?m$ would be possible from $s$, then $s_2$ is a deadlock.
State $s_1$ is not a deadlock however, since it has $m$ at the head of its channel.

If we assume unaltered reachability, then we can derive from Lemma \ref{lem:deadlocksplit} that deadlock freedom is split-preserved.

\begin{theorem}
\label{the:deadlocksplit}
If $S \unre S'$, then deadlock freedom is split-preserved.
\end{theorem}

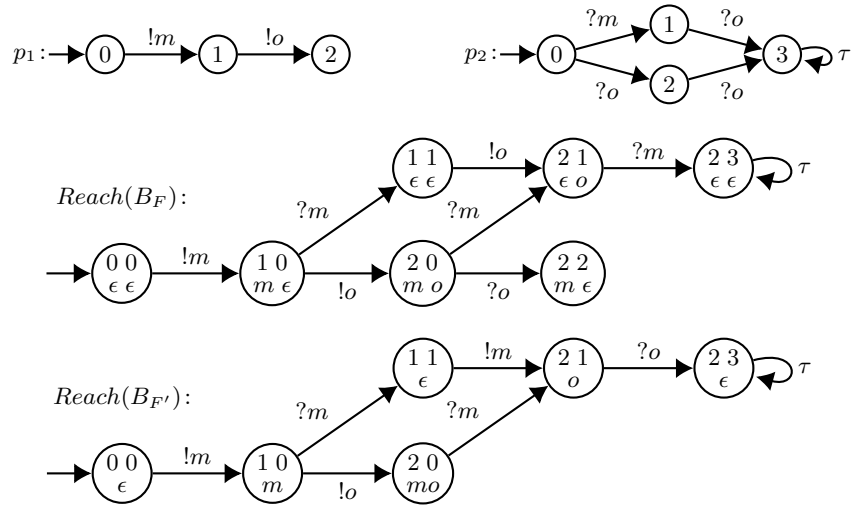
\begin{figure}[b!]
\begin{subfigure}[b]{\textwidth}
\centering
\begin{tikzpicture}[lts]

\node		(name)	at	(-1, -0.02)	{$p_1\!:$};
\coordinate	(init)	at	(-0.75, 0);
\node[state]	(0)	at	(0, 0)		{$0$};
\node[state]	(1)	at	(1.5, 0)	{$1$};
\node[state]	(2)	at	(3, 0)		{$2$};

\path
	(init)	edge									(0)
	(0)		edge				node		{$!m$}	(1)
	(1)		edge				node		{$!o$}	(2);

\node		(name)	at	(5, -0.02)	{$p_2\!:$};
\coordinate	(init)	at	(5.25, 0);
\node[state]	(0)	at	(6, 0)		{$0$};
\node[state]	(1)	at	(7.5, 0.4)	{$1$};
\node[state]	(2)	at	(7.5, -0.4)	{$2$};
\node[state]	(3)	at	(9, 0)		{$3$};

\path
	(init)	edge												(0)
	(0)		edge				node[pos=0.7]		{$?m$}		(1)
			edge				node[swap, pos=0.7]	{$?o$}		(2)
	(1)		edge				node[pos=0.3]		{$?o$}		(3)
	(2)		edge				node[swap, pos=0.3]	{$?o$}		(3)
	(3)		edge[loop right]	node				{$\tau$}	(3);

\end{tikzpicture}
\end{subfigure}\\[1em]
\begin{subfigure}[b]{\textwidth}
\centering
\begin{tikzpicture}[lts]

\node		(name)	at	(0, 1)	{$Reach(B_F)\!:$};
\coordinate	(init)	at	(-1, 0);
\node[state]	(0)	at	(0, 0)		{$0 \; 0$\\[-1mm]$\epsilon \; \epsilon$};
\node[state]	(1)	at	(2, 0)		{$1 \; 0$\\[-1mm]$m \; \epsilon$};
\node[state]	(2)	at	(4, 0)		{$2 \; 0$\\[-1mm]$m \; o$};
\node[state]	(3)	at	(4, 1.4)	{$1 \; 1$\\[-1mm]$\epsilon \; \epsilon$};
\node[state]	(4)	at	(6, 0)		{$2 \; 2$\\[-1mm]$m \; \epsilon$};
\node[state]	(5)	at	(6, 1.4)	{$2 \; 1$\\[-1mm]$\epsilon \; o$};
\node[state]	(6)	at	(8, 1.4)	{$2 \; 3$\\[-1mm]$\epsilon \; \epsilon$};

\path
	(init)	edge											(0)
	(0)		edge				node			{$!m$}		(1)
	(1)		edge				node[swap]		{$!o$}		(2)
			edge				node[pos=0.4]	{$?m$}		(3)
	(2)		edge				node[swap]		{$?o$}		(4)
			edge				node[pos=0.4]	{$?m$}		(5)
	(3)		edge				node			{$!o$}		(5)
	(5)		edge				node			{$?m$}		(6)
	(6)		edge[loop right]	node			{$\tau$}	(6);

\end{tikzpicture}
\end{subfigure}\\[1em]
\begin{subfigure}[b]{\textwidth}
\centering
\begin{tikzpicture}[lts]

\node		(name)	at	(0, 1)	{$Reach(B_{F'})\!:$};
\coordinate	(init)	at	(-1, 0);
\node[state]	(0)	at	(0, 0)		{$0 \; 0$\\[-1mm]$\epsilon$};
\node[state]	(1)	at	(2, 0)		{$1 \; 0$\\[-1mm]$m$};
\node[state]	(2)	at	(4, 0)		{$2 \; 0$\\[-1mm]$mo$};
\node[state]	(3)	at	(4, 1.4)	{$1 \; 1$\\[-1mm]$\epsilon$};
\node[state]	(5)	at	(6, 1.4)	{$2 \; 1$\\[-1mm]$o$};
\node[state]	(6)	at	(8, 1.4)	{$2 \; 3$\\[-1mm]$\epsilon$};

\path
	(init)	edge											(0)
	(0)		edge				node			{$!m$}		(1)
	(1)		edge				node[swap]		{$!o$}		(2)
			edge				node[pos=0.4]	{$?m$}		(3)
	(2)		edge				node[pos=0.4]	{$?m$}		(5)
	(3)		edge				node			{$!m$}		(5)
	(5)		edge				node			{$?o$}		(6)
	(6)		edge[loop right]	node			{$\tau$}	(6);

\end{tikzpicture}
\end{subfigure}

\caption{An example that shows that deadlock freedom is not split-preserved without assuming unaltered reachability ($S \unre S'$), with $F = \langle \{p_1, p_2\}, \{\{m\}, \{o\}\}, \{m, o\} \rangle$ and $F' = \langle \{p_1, p_2\}, \{\{m, o\}\}, \{m, o\} \rangle$.}
\label{fig:deadlock}
\end{figure}

\begin{example}
See Figure \ref{fig:deadlock} for an example that shows why the condition $S \unre S'$ is needed for split-preservation of deadlock freedom.
In $Reach(B_{F'})$, state $2$ of process $p_2$ is not reachable, because the single channel forces $p_2$ to receive $m$ and $o$ in the order that they are sent.
In $Reach(B_F)$, $m$ and $o$ are put in different channels, so $p_2$ is free to choose which it receives first.
This makes $2$ of $p_2$ reachable, which violates unaltered reachability.
The corresponding state in $Reach(B_F)$ is a deadlock, because $p_2$ expects another $o$ which is never supplied.
\end{example}

\subsection{Confluence}

Confluence of two actions indicates a form of independence between them.
Since a FIFO system consists of multiple processes acting mostly independently of each other, confluence in a FIFO system is common.

\begin{definition}
Let $B = \langle S, s_0, L, \trans{}\rangle$ be an LTS.
For $a, b \in L$ and $s \in S$, $a$ and $b$ are \emph{confluent from} $s$, denoted as $\Conf^a_b(s)$, iff for all $t, u \in S$ we have that $(s \trans{a} t \wedge s \trans{b} u) \Rightarrow (\exists_{v \in S} : t \trans{b} v \wedge u \trans{a} v)$.
Note that $\Conf^a_b(s) = \Conf^b_a(s)$.
We say that $a$ and $b$ are \emph{confluent in} $B$ iff $\Conf^a_b(s)$ for all $s \in Reach(S)$.
\end{definition}

Again, we will first look at the preservation on state level.
The relation between confluence and independence is reflected in its preservation: confluence of two actions from a state is preserved when merging channels, if the two actions do not use the same channel.
This is the case when at least one of two actions is $\tau$ and when both actions use different channels in both FIFO systems.
When splitting channels, there is an exception when an input action $a$ is involved that uses a split channel.
If a choice between $a$ and another action exists from a state $s$ after splitting channels, there may be some specialising states in the original LTS from which $a$ is not possible due to the interleaving of channel contents.
This makes confluence trivially true from these states, while confluence may be false from $s$.
We represent this case with the condition $\SC{a}$, which is true iff $a = \;?m \Rightarrow \chan{C}{m} = \chan{C'}{m}$ for some $m \in M$.

\begin{figure}[b!]
\begin{subfigure}{\textwidth}
\centering
\begin{tikzpicture}[lts]

\node		(name)	at	(-1, -0.02)	{$p_1\!:$};
\coordinate	(init)	at	(-0.75, 0);
\node[state]	(0)	at	(0, 0)		{$0$};
\node[state]	(1)	at	(1.3, 0)	{$1$};
\node[state]	(2)	at	(2.6, 0)		{$2$};
\node[state]	(3)	at	(3.9, 0)	{$3$};

\path
	(init)	edge					(0)
	(0)		edge	node	{$!o$}	(1)
	(1)		edge	node	{$!m$}	(2)
	(2)		edge	node	{$!n$}	(3);

\node		(name)	at	(5, -0.02)	{$p_2\!:$};
\coordinate	(init)	at	(5.25, 0);
\node[state]	(0)	at	(6, 0)		{$0$};
\node[state]	(1)	at	(7.3, 0)	{$1$};
\node[state]	(2)	at	(8.6, 0.4)	{$2$};
\node[state]	(3)	at	(8.6, -0.4)	{$3$};

\path
	(init)	edge									(0)
	(0)		edge	node				{$?n$}		(1)
	(1)		edge	node[pos=0.7]		{$?m$}		(2)
			edge	node[swap, pos=0.7]	{$\tau$}	(3);

\end{tikzpicture}
\end{subfigure}\\[1em]
\begin{subfigure}[b]{\textwidth}
\centering
\begin{tikzpicture}[lts]

\node		(name)	at	(-2, 0)		{$Reach(B_F)\!:$};
\coordinate	(init)	at	(-1, 0);
\node[state]	(0)	at	(0, 0)		{$0 \; 0$\\[-1mm]$\epsilon \; \epsilon \; \epsilon$};
\node[state]	(1)	at	(1.6, 0)	{$1 \; 0$\\[-1mm]$\epsilon \; o \; \epsilon$};
\node[state]	(2)	at	(3.2, 0)	{$2 \; 0$\\[-1mm]$m \; o \; \epsilon$};
\node[state]	(3)	at	(4.8, 0)	{$3 \; 0$\\[-1mm]$m \; o \; n$};
\node[state]	(4)	at	(6.4, 0)	{$3 \; 1$\\[-1mm]$m \; o \; \epsilon$};
\node[state]	(5)	at	(8, 0.65)	{$3 \; 2$\\[-1mm]$\epsilon \; o \; \epsilon$};
\node[state]	(6)	at	(8, -0.65)	{$3 \; 3$\\[-1mm]$m \; o \; \epsilon$};

\path
	(init)	edge									(0)
	(0)		edge	node				{$!o$}		(1)
	(1)		edge	node				{$!m$}		(2)
	(2)		edge	node				{$!n$}		(3)
	(3)		edge	node				{$?n$}		(4)
	(4)		edge	node[pos=0.7]		{$?m$}		(5)
			edge	node[swap, pos=0.7]	{$\tau$}	(6);

\end{tikzpicture}
\end{subfigure}\\[1em]
\begin{subfigure}[b]{\textwidth}
\centering
\begin{tikzpicture}[lts]

\node		(name)	at	(-2, 0)		{$Reach(B_{F'})\!:$};
\coordinate	(init)	at	(-1, 0);
\node[state]	(0)	at	(0, 0)		{$0 \; 0$\\[-1mm]$\epsilon \; \epsilon$};
\node[state]	(1)	at	(1.6, 0)	{$1 \; 0$\\[-1mm]$o \; \epsilon$};
\node[state]	(2)	at	(3.2, 0)	{$2 \; 0$\\[-1mm]$om \; \epsilon$};
\node[state]	(3)	at	(4.8, 0)	{$3 \; 0$\\[-1mm]$om \; n$};
\node[state]	(4)	at	(6.4, 0)	{$3 \; 1$\\[-1mm]$om \; \epsilon$};
\node[state]	(5)	at	(8, 0)		{$3 \; 3$\\[-1mm]$om \; \epsilon$};

\path
	(init)	edge						(0)
	(0)		edge	node	{$!o$}		(1)
	(1)		edge	node	{$!m$}		(2)
	(2)		edge	node	{$!n$}		(3)
	(3)		edge	node	{$?n$}		(4)
	(4)		edge	node	{$\tau$}	(5);

\end{tikzpicture}
\end{subfigure}

\caption{An example that shows that without condition $\SC{a}$ confluence is not split-preserved, with $F = \langle \{p_1, p_2\}, \{\{m\}, \{o\}, \{n\}\}, \{m, o, n\} \rangle$ and $F' = \langle \{p_1, p_2\}, \{\{m, o\}, \{n\}\}, \{m, o, n\} \rangle$.}
\label{fig:confsplit}
\end{figure}
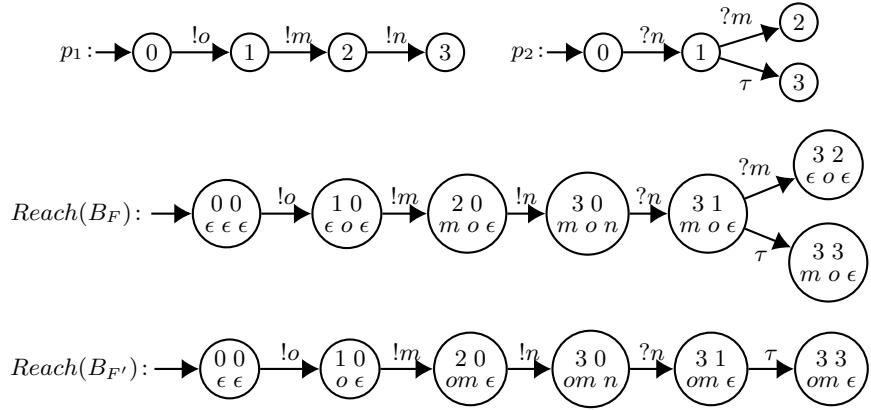

\begin{lemma}
\label{lem:conftau}
Let $s \in S$, $s' \in S'$ and $a \in L$.
Assume that $s \corr s'$.
Then $\Conf^\tau_a(s) \Rightarrow \Conf^\tau_a(s')$ and if $\SC{a}$, then $\Conf^\tau_a(s) \Leftarrow \Conf^\tau_a(s')$.
\end{lemma}

\begin{lemma}
\label{lem:confoc}
Let $s \in S$, $s' \in S'$ and $m, o \in M$.
Assume that $s \corr s'$ and $m \not\simeq_{C'} o$.
Let $a \in \{?m, !m\}$ and $b \in \{?o, !o\}$.
Then $\Conf^a_b(s) \Rightarrow \Conf^a_b(s')$ and if $\SC{a}$ and $\SC{b}$, then $\Conf^a_b(s) \Leftarrow \Conf^a_b(s')$.
\end{lemma}

\begin{example}
See Figure \ref{fig:confsplit} for an example why condition $\SC{a}$ is necessary for the split-preservation of confluence.
In $Reach(B_F)$ confluence of $?m$ and $\tau$ is not met, because there is a choice between the two that does not result in a confluence diamond.
In $Reach(B_{F'})$ confluence of $?m$ and $\tau$ is trivially met because the choice between the two is never possible.
Compared to $Reach(B_F)$, the choice was made impossible in $Reach(B_{F'})$ because the channels for $m$ and $o$ have now merged.
Because $o$ is sent before $m$, $p_2$ is forced in $Reach(B_{F'})$ to receive $o$ first, but it never does.
Note that $\SC{?m}$ is not met.
\end{example}

If both actions $a$ and $b$ use the same channel in both FIFO systems, there is an edge case where confluence is not preserved when merging channels, namely when both are the exact same input action.
In this case, two messages $m$ are required at the head of the channel of $m$ to create the confluence diamond.
However, if the channel of $m$ is merged with another channel, there are specialised states with an interleaving of channel contents without both messages $m$ in front.
We represent this case with $a \equiv_? b$ for actions $a, b \in L$, which is true iff $a = \;?m = b$ for some $m \in M$.

\begin{lemma}
\label{lem:confsc}
Let $s \in S$, $s' \in S'$ and $m, o \in M$.
Assume that $s \corr s'$ and $m \simeq_C o$.
Let $a \in \{?m, !m\}$ and $b \in \{?o, !o\}$.
If not $a \equiv_? b$, then $\Conf^a_b(s) \Rightarrow \Conf^a_b(s')$ and if $\SC{a}$ and $\SC{b}$, then $\Conf^a_b(s) \Leftarrow \Conf^a_b(s')$.
\end{lemma}

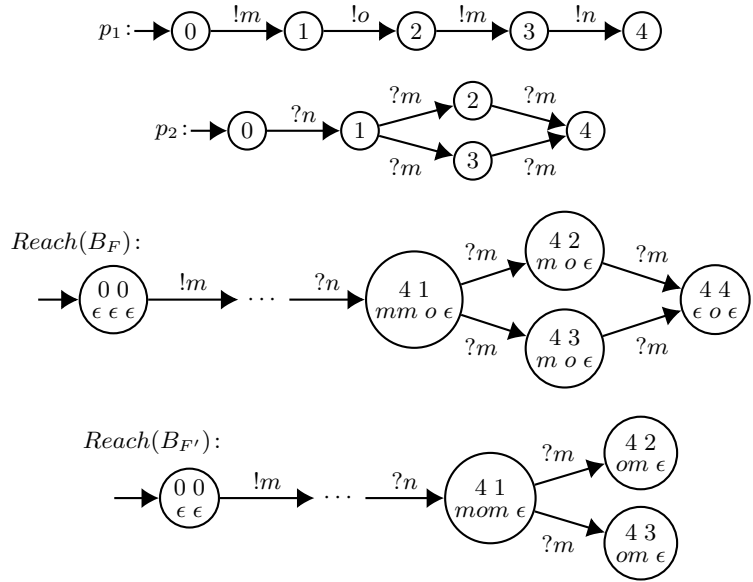
\begin{figure}[!t]
\begin{subfigure}[b]{\textwidth}
\centering
\begin{tikzpicture}[lts]

\node		(name)	at	(-1, -0.02)	{$p_1\!:$};
\coordinate	(init)	at	(-0.75, 0);
\node[state]	(0)	at	(0, 0)		{$0$};
\node[state]	(1)	at	(1.5, 0)	{$1$};
\node[state]	(2)	at	(3, 0)		{$2$};
\node[state]	(3)	at	(4.5, 0)	{$3$};
\node[state]	(4)	at	(6, 0)		{$4$};

\path
	(init)	edge					(0)
	(0)		edge	node	{$!m$}	(1)
	(1)		edge	node	{$!o$}	(2)
	(2)		edge	node	{$!m$}	(3)
	(3)		edge	node	{$!n$}	(4);

\end{tikzpicture}
\end{subfigure}\\[1em]
\begin{subfigure}[b]{\textwidth}
\centering
\begin{tikzpicture}[lts]

\node		(name)	at	(-1, -0.02)	{$p_2\!:$};
\coordinate	(init)	at	(-0.75, 0);
\node[state]	(0)	at	(0, 0)		{$0$};
\node[state]	(1)	at	(1.5, 0)	{$1$};
\node[state]	(2)	at	(3, 0.4)	{$2$};
\node[state]	(3)	at	(3, -0.4)	{$3$};
\node[state]	(4)	at	(4.5, 0)	{$4$};

\path
	(init)	edge								(0)
	(0)		edge	node				{$?n$}	(1)
	(1)		edge	node[pos=0.7]		{$?m$}	(2)
			edge	node[swap, pos=0.7]	{$?m$}	(3)
	(2)		edge	node[pos=0.3]		{$?m$}	(4)
	(3)		edge	node[swap, pos=0.3]	{$?m$}	(4);

\end{tikzpicture}
\end{subfigure}\\[1em]
\begin{subfigure}[b]{\textwidth}
\centering
\begin{tikzpicture}[lts]

\node		(name)	at	(-0.5, 0.75)	{$Reach(B_F)\!:$};
\coordinate	(init)	at	(-1, 0);
\node[state]	(0)	at	(0, 0)		{$0 \; 0$\\[-1mm]$\epsilon \; \epsilon \; \epsilon$};
\node			(1)	at	(2, 0)		{$\cdots$};
\node[state]	(2)	at	(4, 0)		{$4 \; 1$\\[-1mm]$mm \; o \; \epsilon$};
\node[state]	(3)	at	(6, 0.65)	{$4 \; 2$\\[-1mm]$m \; o \; \epsilon$};
\node[state]	(4)	at	(6, -0.65)	{$4 \; 3$\\[-1mm]$m \; o \; \epsilon$};
\node[state]	(5)	at	(8, 0)		{$4 \; 4$\\[-1mm]$\epsilon \; o \; \epsilon$};

\path
	(init)	edge								(0)
	(0)		edge	node				{$!m$}	(1)
	(1)		edge	node				{$?n$}	(2)
	(2)		edge	node[pos=0.7]		{$?m$}	(3)
			edge	node[swap,pos=0.7]	{$?m$}	(4)
	(3)		edge	node[pos=0.3]		{$?m$}	(5)
	(4)		edge	node[swap, pos=0.3]	{$?m$}	(5);

\end{tikzpicture}
\end{subfigure}\\[1em]
\begin{subfigure}[b]{\textwidth}
\centering
\begin{tikzpicture}[lts]

\node		(name)	at	(-0.5, 0.75)	{$Reach(B_{F'})\!:$};
\coordinate	(init)	at	(-1, 0);
\node[state]	(0)	at	(0, 0)		{$0 \; 0$\\[-1mm]$\epsilon \; \epsilon$};
\node			(1)	at	(2, 0)		{$\cdots$};
\node[state]	(2)	at	(4, 0)		{$4 \; 1$\\[-1mm]$mom \; \epsilon$};
\node[state]	(3)	at	(6, 0.6)	{$4 \; 2$\\[-1mm]$om \; \epsilon$};
\node[state]	(4)	at	(6, -0.6)	{$4 \; 3$\\[-1mm]$om \; \epsilon$};

\path
	(init)	edge								(0)
	(0)		edge	node				{$!m$}	(1)
	(1)		edge	node				{$?n$}	(2)
	(2)		edge	node[pos=0.7]		{$?m$}	(3)
			edge	node[swap, pos=0.7]	{$?m$}	(4);

\end{tikzpicture}
\end{subfigure}

\caption{An example that shows that without condition not $a \equiv_? b$ confluence is not merge-preserved, with $F = \langle \{p_1, p_2\}, \{\{m\}, \{o\}, \{n\}\}, \{m, o, n\} \rangle$ and $F' = \langle \{p_1, p_2\}, \{\{m, o\}, \{n\}\}, \{m, o, n\} \rangle$.}
\label{fig:sameconf}
\end{figure}

\begin{example}
See Figure \ref{fig:sameconf} for an example why condition not $a \equiv_? b$ is necessary for merge-preservation of confluence.
In $Reach(B_F)$ confluence of $?m$ and $?m$ is met, because the only choice between $?m$ and $?m$ results in a confluence diamond.
In $Reach(B_{F'})$ confluence of $?m$ and $?m$ is not met, because there is a choice between $?m$ and $?m$ that does not result in a confluence diamond.
This is because $p_2$ first needs to receive the $o$ before it can receive the second $m$.
In $Reach(B_F)$ this was not an issue, because $m$ and $o$ both had their own channels.
Note that $?m \equiv_? ?m$.
\end{example}

If both actions use channels that are distinct in $F$ but equal in $F'$, confluence of two actions is merge-preserved if at least one of the actions is an input action.
In case one action is an input action and the other an output action, merge-preservation follows from the fact that the actions touch different ends of the channel and are therefore in some sense independent.
In case both actions are input actions, the choice between the two actions is not possible from any state in $B_{F'}$, since they use the same channel.
This makes their confluence hold trivially, from which merge-preservation trivially follows.
Confluence is not split-preserved in these cases for the same reason as for the example in Figure \ref{fig:confsplit} (for instance, replace $\tau$ with $?o$).

In case both actions are output actions, confluence is only split-preserved.
This is because from any state $s'$ in $B_{F'}$, the two different orders of these actions produce different orders of channel contents, since both actions use the same channel.
This implies that confluence cannot hold from $s'$, and therefore it is trivially preserved when splitting channels.
Confluence is still possible in $B_F$, so confluence is not necessarily preserved when merging channels.
For actions $a, b \in L$, we formally represent this last case with 	$\AC{a}{b}$, which is true iff $a =\; !m$, $b =\; !o$, $m \not\simeq_C o$ and $m \simeq_{C'} o$ for some $m, o \in M$.

\begin{lemma}
\label{lem:confccmerge}
Let $s \in S$, $s' \in S'$ and $m, o \in M$.
Assume that $s \corr s'$, $m \not\simeq_C o$ and $m \simeq_{C'} o$.
Let $a \in \{?m, !m\}$ and $b \in \{?o, !o\}$.
If not $\AC{a}{b}$, then $\Conf^{a}_{b}(s) \Rightarrow \Conf^{a}_{b}(s')$.
\end{lemma}

\begin{lemma}
\label{lem:confccsplit}
Let $s \in S$, $s' \in S'$ and $a, b \in L$
Assume that $s \corr s'$.
If $\AC{a}{b}$, then $\Conf^a_b(s) \Leftarrow \Conf^a_b(s')$.
\end{lemma}

Lifting these state-based results to confluence in the LTSs of FIFO systems, we can derive the following theorems:

\begin{theorem}
\label{the:confmerge}
Let $a, b \in L$.
If not $\AC{a}{b}$ and not $a \equiv_? b$, then confluence of $a$ and $b$ is merge-preserved.
\end{theorem}

\begin{theorem}
\label{the:confsplit}
Let $a, b \in L$.
If $\SC{a}$, $\SC{b}$ and $S \unre S'$, then confluence of $a$ and $b$ is split-preserved.
\end{theorem}

\subsection{Summary of results}

The results of this section are summarised in the table below.
Remember that $\IL(F, F') = \{\tau\} \cup \{?m, !m\ |\ m \in M \wedge \chan{C}{m} = \chan{C'}{m}\}$, that $S \unre S'$ iff $\forall_{s \in Reach(S)} : \exists_{s' \in Reach(S')} : s \corr s'$, that $\SC{a}$ iff $a = \;?m \Rightarrow \chan{C}{m} = \chan{C'}{m}$ for some $m \in M$, that $a \equiv_? b$ iff $a = \;?m = b$ for some $m \in M$, and that $\AC{a}{b}$ iff $a =\; !m$, $b =\; !o$, $m \not\simeq_C o$ and $m \simeq_{C'} o$ for some $m, o \in M$.\\[1em]
\begin{tabular}{p{0.27\textwidth}||p{0.33\textwidth}|p{0.33\textwidth}}
 & Merge-preserved & Split-preserved\\
\hline\hline
$L'$-reachability of $\kappa$ & If $L' = \IL(F, F')$ (Th. \ref{the:reachmerge}) & If $L' = L$ (Th. \ref{the:reachsplit})\\
\hline
Deadlock freedom & No & If $S \unre S'$ (Th. \ref{the:deadlocksplit})\\
\hline
Confluence of $a$ and $b$ & If not $\AC{a}{b}$ and not $a \equiv_? b$ (Th. \ref{the:confmerge}) & If $\SC{a}$, $\SC{b}$ and $S \unre S'$ (Th. \ref{the:confsplit})
\end{tabular}

\section{Conclusion}
\label{sec:conclusion}


We have studied asynchronously communicating systems and their channel layouts by modelling them as FIFO systems and ordered them based on whether one can be created from the other by merging channels.
We have shown that the LTS that describes the behaviour of a split FIFO system simulates the LTS of the original system.
As a consequence of this, safety properties are merge-preserved.
We have also identified conditions under which reachability, deadlock freedom and confluence are preserved when changing the channel layout.

For most conditions that are required for a property to be preserved, their truth can be derived easily.
An exception of this is the unaltered reachability assumption, for which it should be investigated how feasible it is to check them.
It is also the question how likely the conditions are met in practice, given that some are rather strict.
Using more detailed information from the processes of the FIFO systems could lead to less strict conditions, but they can be more difficult to check.
Another option would be to find whether sufficient property-specific conditions exist.

The properties mentioned in this paper are of course not the only interesting properties one could want to be preserved.
Other options would be preservation of maximum queue length, of eventual termination, of (lack of) starvation and of behavioural equivalence between two systems.

\section*{Acknowledgements}

We thank the reviewers for their helpful feedback.

\bibliographystyle{alpha}
\bibliography{bib}

\newcommand{\etalchar}[1]{$^{#1}$}
\begin{thebibliography}{CHNQ19}

\bibitem[BFS20]{DBLP:conf/concur/BolligFS20}
Benedikt Bollig, Alain Finkel, and Amrita Suresh.
\newblock Bounded reachability problems are decidable in {FIFO} machines.
\newblock In {\em {CONCUR}}, volume 171 of {\em LIPIcs}, pages 49:1--49:17.
  Schloss Dagstuhl - Leibniz-Zentrum f{\"{u}}r Informatik, 2020.

\bibitem[CHNQ19]{DBLP:conf/fm/ChevrouH0Q19}
Florent Chevrou, Aur{\'{e}}lie Hurault, Shin Nakajima, and Philippe
  Qu{\'{e}}innec.
\newblock A map of asynchronous communication models.
\newblock In {\em {FM} Workshops {(2)}}, volume 12233 of {\em Lecture Notes in
  Computer Science}, pages 307--322. Springer, 2019.

\bibitem[CHQ16]{DBLP:journals/fac/ChevrouHQ16}
Florent Chevrou, Aur{\'{e}}lie Hurault, and Philippe Qu{\'{e}}innec.
\newblock On the diversity of asynchronous communication.
\newblock {\em Formal Aspects Comput.}, 28(5):847--879, 2016.

\bibitem[DS12]{DBLP:journals/fac/DerrickS12}
John Derrick and Graeme Smith.
\newblock Temporal-logic property preservation under {Z} refinement.
\newblock {\em Formal Aspects Comput.}, 24(3):393--416, 2012.

\bibitem[EMR02]{DBLP:journals/scp/EngelsMR02}
Andr{\'{e}} Engels, Sjouke Mauw, and Michel~A. Reniers.
\newblock A hierarchy of communication models for message sequence charts.
\newblock {\em Sci. Comput. Program.}, 44(3):253--292, 2002.

\bibitem[FP20]{DBLP:journals/lmcs/FinkelP20}
Alain Finkel and M.~Praveen.
\newblock Verification of flat {FIFO} systems.
\newblock {\em Log. Methods Comput. Sci.}, 16(4), 2020.

\bibitem[HJMS03]{DBLP:conf/birthday/HenzingerJMS03}
Thomas~A. Henzinger, Ranjit Jhala, Rupak Majumdar, and Marco A.~A. Sanvido.
\newblock Extreme model checking.
\newblock In {\em Verification: Theory and Practice}, volume 2772 of {\em
  Lecture Notes in Computer Science}, pages 332--358. Springer, 2003.

\bibitem[HVG03]{DBLP:conf/memocode/HuangVG03}
Jinfeng Huang, Jeroen Voeten, and Marc Geilen.
\newblock Real-time property preservation in approximations of timed systems.
\newblock In {\em {MEMOCODE}}, pages 163--171. {IEEE} Computer Society, 2003.

\bibitem[Lam78]{DBLP:journals/cacm/Lamport78}
Leslie Lamport.
\newblock Time, clocks, and the ordering of events in a distributed system.
\newblock {\em Commun. {ACM}}, 21(7):558--565, 1978.

\bibitem[LGS{\etalchar{+}}95]{DBLP:journals/fmsd/LoiseauxGSBB95}
Claire Loiseaux, Susanne Graf, Joseph Sifakis, Ahmed Bouajjani, and Saddek
  Bensalem.
\newblock Property preserving abstractions for the verification of concurrent
  systems.
\newblock {\em Formal Methods Syst. Des.}, 6(1):11--44, 1995.

\bibitem[SS94]{DBLP:conf/cav/SokolskyS94}
Oleg Sokolsky and Scott~A. Smolka.
\newblock Incremental model checking in the modal mu-calculus.
\newblock In {\em {CAV}}, volume 818 of {\em Lecture Notes in Computer
  Science}, pages 351--363. Springer, 1994.

\bibitem[WE13]{DBLP:conf/tacas/WijsE13}
Anton Wijs and Luc Engelen.
\newblock Efficient property preservation checking of model refinements.
\newblock In {\em {TACAS}}, volume 7795 of {\em Lecture Notes in Computer
  Science}, pages 565--579. Springer, 2013.

\bibitem[Weh00]{DBLP:conf/fmoods/Wehrheim00}
Heike Wehrheim.
\newblock Behavioural subtyping and property preservation.
\newblock In {\em {FMOODS}}, volume 177 of {\em {IFIP} Conference Proceedings},
  pages 213--231. Kluwer, 2000.

\bibitem[XL21]{DBLP:journals/ieeejas/XiaL21}
Chuanliang Xia and Chengdong Li.
\newblock Property preservation of petri synthesis net based representation for
  embedded systems.
\newblock {\em {IEEE} {CAA} J. Autom. Sinica}, 8(4):905--915, 2021.

\end{thebibliography}

\appendix

\section{Proofs}

\noindent\textbf{Lemma \ref{lem:statecorr}.}
\textit{
Let $F = \langle P, C, M \rangle$ and $F' = \langle P, C', M \rangle$ be FIFO systems such that $F \cgtc F'$.
Let $B_F = \langle S, s_0, L, \trans{} \rangle$ and $B_{F'} = \langle S', s_0', L, \ptrans{} \rangle$.
Then $\forall_{s' \in S'} : \exists_{s \in S} : s \corr s'$ and $\forall_{s \in S} : \exists_{s' \in S'} : s \corr s'$.
}
\begin{proof}
We first prove $\forall_{s' \in S'} : \exists_{s \in S} : s \corr s'$.
Pick some $s' \in S'$.
Let $s' = (\kappa, \zeta')$ where $\kappa \in \mathbf{P}$ and $\zeta' \in \mathbf{C}'$.
Using Definition \ref{def:statecorr}, what is left to prove is whether there exists a state $s = (\kappa, \zeta) \in S$ for $\zeta \in \mathbf{C}$ such that $\zeta' \in \mathbf{C}_\zeta'$.
To pick a $\zeta$, we first define the projection function $\pi_{M'}(w)$ for $M' \subseteq M$ and $w \in M^*$, such that $\pi_{M'}(\epsilon) = \epsilon$, $\pi_{M'}(w) = \pi_{M'}(tl(w))$ if $hd(w) \not\in M'$, and $\pi_{M'}(w) = hd(w) \concat \pi_{M'}(tl(w))$ if $hd(w) \in M'$.
Then we pick $\zeta$ such that $\zeta(c) = \pi_c(\zeta'(c'))$ for all $c \in C$ with $c' \in C'$ such that $c \subseteq c'$.
By Definition \ref{def:cgtc} we know that exactly one such $c'$ exists for each $c$.
Also, by the definition of $S$ (Definition \ref{def:FIFO2LTS}), we know that such a $\zeta$ exists, since $\zeta \in \mathbf{C}$.
By Definition \ref{def:chancorr}, what is left to show is that $\zeta'(c') \in \biginterl\{\zeta(c)\ |\ c \in C \wedge c \subseteq c'\}$ for all $c' \in C'$.
We prove this by induction on the length of $\zeta'(c')$.
If $\zeta'(c') = \epsilon$, then from $\zeta(c) = \pi_c(\zeta'(c'))$ it follows that $\zeta(c) = \epsilon$ for all $c \in C$ such that $c \subseteq c'$, thus $\biginterl\{\zeta(c)\ |\ c \in C \wedge c \subseteq c'\} = \{\epsilon\}$.
In case $\zeta'(c') = m \concat w$ for some $m \in M$ and $w \in M^*$, we know that$\chan{C}{m} \subseteq c'$ and by definition of $\pi_c$ that there is some $v \in M^*$ such that $\zeta(\chan{C}{m}) = m \concat v$.
Following the definition of $\biginterl$, we know that $(m \concat \biginterl((\{\zeta(c)\ |\ c \in C \wedge c \subseteq c'\} \setminus \{m \concat v\}) \cup \{v\})) \subseteq (\biginterl\{\zeta(c)\ |\ c \in C \wedge c \subseteq c'\})$.
By the induction hypothesis, we know that $w \in \biginterl((\{\zeta(c)\ |\ c \in C \wedge c \subseteq c'\} \setminus \{m \concat v\}) \cup \{v\})$, thus $m \concat w \in m \concat \biginterl((\{\zeta(c)\ |\ c \in C \wedge c \subseteq c'\} \setminus \{m \concat v\}) \cup \{v\})$, from which we can conclude that $\zeta'(c') \in \biginterl\{\zeta(c)\ |\ c \in C \wedge c \subseteq c'\}$.

Next we prove that $\forall_{s \in S} : \exists_{s' \in S'} : s \corr s'$.
Pick some $s \in S$.
Let $s = (\kappa, \zeta)$ where $\kappa \in \mathbf{P}$ and $\zeta \in \mathbf{C}$.
Using Definition \ref{def:statecorr}, what is left to prove is whether there exists a state $s' = (\kappa, \zeta') \in S'$ for $\zeta' \in \mathbf{C}'$ such that $\zeta' \in \mathbf{C}_\zeta'$.
By Definition \ref{def:chancorr}, it is not hard to see that $\mathbf{C}_\zeta' \neq \emptyset$, so a $\zeta'$ exists.
Also, from this definition we know that $\zeta' \in \mathbf{C}'$.
Then by the definition of $S'$ (Definition \ref{def:FIFO2LTS}), we can conclude that $s'$ exists.
\qed
\end{proof}

\noindent\textbf{Lemma \ref{lem:transcorrtau}.}
\textit{
Let $F = \langle P, C, M \rangle$ and $F' = \langle P, C', M \rangle$ be FIFO systems such that $F \cgtc F'$.
Let $B_F = \langle S, s_0, L, \trans{} \rangle$ and $B_{F'} = \langle S', s_0', L, \ptrans{} \rangle$.
Then for any $\kappa_1, \kappa_2 \in \mathbf{P}$, $\zeta \in \mathbf{C}$ and $\zeta' \in \mathbf{C}_{\zeta}'$, it holds that $(\kappa_1, \zeta) \trans{\tau} (\kappa_2, \zeta)$ iff $(\kappa_1, \zeta') \ptrans{\tau} (\kappa_2, \zeta')$.
}
\begin{proof}
The bi-implication is proven in both directions.
\begin{itemize}
\item[$\Rightarrow$:]
Let $(\kappa_1, \zeta) \trans{\tau} (\kappa_2, \zeta)$.
According to the definition of $\trans{}$, a $\tau$-step is only possible if $\kappa_2 = \kappa_1[p \mapsto q]$ for some $p \in P$ and $q \in Q_p$, so we can assume this.
If we then apply the definition of $\trans{}$ we get that $\kappa_1(p) \Dtrans[p]{\tau} q$.
Then applying the definition of $\ptrans{}$, it follows that $(\kappa_1, \zeta') \ptrans{\tau} (\kappa_1[p \mapsto q], \zeta')$.
Since $\kappa_2 = \kappa_1[p \mapsto q]$, we can conclude that $(\kappa_1, \zeta') \ptrans{\tau} (\kappa_2, \zeta')$.
\item[$\Leftarrow$:]
Let $(\kappa_1, \zeta') \ptrans{\tau} (\kappa_2, \zeta')$.
According to the definition of $\ptrans{}$, a $\tau$-step is only possible if $\kappa_2 = \kappa_1[p \mapsto q]$ for some $p \in P$ and $q \in Q_p$, so we can assume this.
If we then apply the definition of $\ptrans{}$ we get that $\kappa_1(p) \Dtrans[p]{\tau} q$.
Then applying the definition of $\trans{}$, it follows that $(\kappa_1, \zeta) \trans{\tau} (\kappa_1[p \mapsto q], \zeta)$.
Since $\kappa_2 = \kappa_1[p \mapsto q]$ we can conclude that $(\kappa_1, \zeta) \trans{\tau} (\kappa_2, \zeta)$.
\end{itemize}
\qed
\end{proof}

\noindent\textbf{Lemma \ref{lem:transcorrrecnomerge}.}
\textit{
Let $F = \langle P, C, M \rangle$ and $F' = \langle P, C', M \rangle$ be FIFO systems such that $F \cgtc F'$.
Let $B_F = \langle S, s_0, L, \trans{} \rangle$ and $B_{F'} = \langle S', s_0', L, \ptrans{} \rangle$.
Then for any $\kappa_1, \kappa_2 \in \mathbf{P}$, $\zeta \in \mathbf{C}$, $\zeta' \in \mathbf{C}_{\zeta}'$ and $m \in M$, with $c = \chan{C}{m}$ and $c' = \chan{C'}{m}$ it holds that if $c = c'$, then $(\kappa_1, \zeta) \trans{?m} (\kappa_2, \zeta[c \mapsto tl(\zeta(c))])$ iff $(\kappa_1, \zeta') \ptrans{?m} (\kappa_2, \zeta'[c' \mapsto tl(\zeta'(c'))])$.
}
\begin{proof}
The bi-implication is proven in both directions.
Let $c = c'$.
\begin{itemize}
\item[$\Rightarrow$:]
Let $(\kappa_1, \zeta) \trans{?m} (\kappa_2, \zeta[c \mapsto tl(\zeta(c))])$.
According to the definition of $\trans{}$, a $?m$-step is only possible if $\kappa_2 = \kappa_1[p \mapsto q]$ for some $p \in P$ and $q \in Q_p$, so we can assume this.
If we then apply the definition of $\trans{}$ we get that $\kappa_1(p) \Dtrans[p]{?m} q \wedge hd(\zeta(c)) = m$.
Using Definition \ref{def:statecorr} and $c = c'$, we can derive from $hd(\zeta(c)) = m$ that $hd(\zeta'(c')) = m$.
Then applying the definition of $\ptrans{}$, it follows that $(\kappa_1, \zeta') \ptrans{?m} (\kappa_1[p \mapsto q], \zeta'[c' \mapsto tl(\zeta'(c'))])$.
Since $\kappa_2 = \kappa_1[p \mapsto q]$ we can conclude that $(\kappa_1, \zeta') \ptrans{?m} (\kappa_2, \zeta'[c' \mapsto tl(\zeta'(c'))])$.
\item[$\Leftarrow$:]
Let $(\kappa_1, \zeta') \ptrans{?m} (\kappa_2, \zeta'[c' \mapsto tl(\zeta'(c'))])$.
According to the definition of $\ptrans{}$, a $?m$-step is only possible if $\kappa_2 = \kappa_1[p \mapsto q]$ for some $p \in P$ and $q \in Q_p$, so we can assume this.
If we then apply the definition of $\ptrans{}$ we get that $\kappa_1(p) \Dtrans[p]{?m} q \wedge hd(\zeta'(c')) = m$.
Using Definition \ref{def:statecorr}, we can derive from $hd(\zeta'(c')) = m$ that $hd(\zeta(c)) = m$.
Then applying the definition of $\trans{}$, it follows that $(\kappa_1, \zeta) \trans{?m} (\kappa_1[p \mapsto q], \zeta[c \mapsto tl(\zeta(c))])$.
Since $\kappa_2 = \kappa_1[p \mapsto q]$, we can conclude that $(\kappa_1, \zeta) \trans{?m} (\kappa_2, \zeta[c \mapsto tl(\zeta(c))])$.
\end{itemize}
\qed
\end{proof}

\noindent\textbf{Lemma \ref{lem:transcorrrecmerge}.}
\textit{
Let $F = \langle P, C, M \rangle$ and $F' = \langle P, C', M \rangle$ be FIFO systems such that $F \cgtc F'$.
Let $B_F = \langle S, s_0, L, \trans{} \rangle$ and $B_{F'} = \langle S', s_0', L, \ptrans{} \rangle$.
Then for any $\kappa_1, \kappa_2 \in \mathbf{P}$, $\zeta \in \mathbf{C}$, $\zeta' \in \mathbf{C}_{\zeta}'$ and $m \in M$, with $c = \chan{C}{m}$ and $c' = \chan{C'}{m}$ it holds that if $c \neq c'$, then $(\kappa_1, \zeta) \trans{?m} (\kappa_2, \zeta[c \mapsto tl(\zeta(c))]) \wedge hd(\zeta'(c')) = m$ iff $(\kappa_1, \zeta') \ptrans{?m} (\kappa_2, \zeta'[c' \mapsto tl(\zeta'(c'))])$.
}
\begin{proof}
The bi-implication is proven in both directions.
Let $c \neq c'$.
\begin{itemize}
\item[$\Rightarrow$:]
Same as the proof for Lemma \ref{lem:transcorrrecnomerge}, except that we use $hd(\zeta'(c')) = m$ from the assumptions since in this case it cannot be derived from $hd(\zeta(c)) = m$.
\item[$\Leftarrow$:]
Same as the proof for Lemma \ref{lem:transcorrrecnomerge}, except that we take $hd(\zeta'(c')) = m$ along to the conclusion.
\end{itemize}
\qed
\end{proof}

\noindent\textbf{Lemma \ref{lem:transcorrsend}.}
\textit{
Let $F = \langle P, C, M \rangle$ and $F' = \langle P, C', M \rangle$ be FIFO systems such that $F \cgtc F'$.
Let $B_F = \langle S, s_0, L, \trans{} \rangle$ and $B_{F'} = \langle S', s_0', L, \ptrans{} \rangle$.
Then for any $\kappa_1, \kappa_2 \in \mathbf{P}$, $\zeta \in \mathbf{C}$, $\zeta' \in \mathbf{C}_{\zeta}'$ and $m \in M$, with $c = \chan{C}{m}$ and $c' = \chan{C'}{m}$ it holds that $(\kappa_1, \zeta) \trans{!m} (\kappa_2, \zeta[c \mapsto \zeta(c) \concat m])$ iff $(\kappa_1, \zeta') \ptrans{!m} (\kappa_2, \zeta'[c' \mapsto \zeta'(c') \concat m])$.
}
\begin{proof}
The bi-implication is proven in both directions.
\begin{itemize}
\item[$\Rightarrow$:]
Let $(\kappa_1, \zeta) \trans{!m} (\kappa_2, \zeta[c \mapsto \zeta(c) \concat m])$.
According to the definition of $\trans{}$, a $!m$-step is only possible if $\kappa_2 = \kappa_1[p \mapsto q]$ for some $p \in P$ and $q \in Q_p$, so we can assume this.
If we then apply the definition of $\trans{}$ we get that $\kappa_1(p) \Dtrans[p]{!m} q$.
Then applying the definition of $\ptrans{}$, it follows that $(\kappa_1, \zeta') \ptrans{!m} (\kappa_1[p \mapsto q], \zeta'[c' \mapsto \zeta'(c') \concat m])$.
Since $\kappa_2 = \kappa_1[p \mapsto q]$, we can conclude that $(\kappa_1, \zeta') \ptrans{!m} (\kappa_2, \zeta'[c' \mapsto \zeta'(c') \concat m])$.
\item[$\Leftarrow$:]
Let $(\kappa_1, \zeta') \ptrans{!m} (\kappa_2, \zeta'[c' \mapsto \zeta'(c') \concat m])$.
According to the definition of $\ptrans{}$, a $!m$-step is only possible if $\kappa_2 = \kappa_1[p \mapsto q]$ for some $p \in P$ and $q \in Q_p$, so we can assume this.
If we then apply the definition of $\ptrans{}$ we get that $\kappa_1(p) \Dtrans[p]{!m} q$.
Then applying the definition of $\trans{}$, it follows that $(\kappa_1, \zeta) \trans{!m} (\kappa_1[p \mapsto q], \zeta[c \mapsto \zeta(c) \concat m])$.
Since $\kappa_2 = \kappa_1[p \mapsto q]$, we can conclude that $(\kappa_1, \zeta) \trans{!m} (\kappa_2, \zeta[c \mapsto \zeta(c) \concat m])$.
\end{itemize}
\qed
\end{proof}

\noindent\textbf{Lemma \ref{lem:corrsim}.}
\textit{
Let $F = \langle P, C, M \rangle$ and $F' = \langle P, C', M \rangle$ be FIFO systems such that $F \cgtc F'$.
Let $B_F = \langle S, s_0, L, \trans{} \rangle$ and $B_{F'} = \langle S', s_0', L, \ptrans{} \rangle$.
Then $\corr^{-1}$ is a simulation relation.
}
\begin{proof}
Pick some $(s', s) \in \corr^{-1}$ and some $a \in L$ and $t' \in S'$ such that $s' \ptrans{a} t'$.
From $s' \corr^{-1} s$ we know that $s \corr s'$, so we can use Lemma \ref{lem:transcorrtau}-\ref{lem:transcorrsend} to show that there exists a $t \in S$ such that $s \trans{a} t$.
With the same lemmas and Definition \ref{def:statecorr} we can derive that $t \corr t'$.
From this it follows that $t' \corr^{-1} t$, from which we can conclude that $\corr^{-1}$ is a simulation relation.

What is left to show is that Lemma \ref{lem:transcorrtau}-\ref{lem:transcorrsend} cover all transitions.
For $\tau$-transitions, Lemma \ref{lem:transcorrtau} only requires that $\zeta$ is the same in the source and target state, which is also required by Definition \ref{def:FIFO2LTS}, so all $\tau$-transitions are covered.
For $?m$-transitions the condition $c = c'$ in Lemma \ref{lem:transcorrrecnomerge} and the condition $c \neq c'$ in Lemma \ref{lem:transcorrrecmerge} cancel each other out.
Both lemmas require that $\zeta$ is updated with $c \mapsto tl(\zeta(c))$ (respectively $c' \mapsto tl(\zeta'(c'))$) in the target state, which is also required by Definition \ref{def:FIFO2LTS}.
Lemma \ref{lem:transcorrrecmerge} additionally requires that $hd(\zeta'(c')) = m$ for the right-hand side to be true, which is also required by Definition \ref{def:FIFO2LTS} (see the right-hand side), so all $?m$-transitions are covered.
For $!m$-transitions, Lemma \ref{lem:transcorrsend} only requires that the channel is updated with $c \mapsto \zeta(c) \concat m$ (respectively $c' \mapsto \zeta'(c') \concat m$, which is also required by Definition \ref{def:FIFO2LTS}, so all $!m$-transitions are covered.
\qed
\end{proof}

\noindent\textbf{Theorem \ref{the:simulation}.}
\textit{
Let $F$ and $F'$ be FIFO systems such that $F \cgtc F'$.
Then $B_F$ simulates $B_{F'}$.
}
\begin{proof}
Let $B_F = \langle S, s_0, L, \trans{} \rangle$ and $B_{F'} = \langle S', s_0', L, \ptrans{} \rangle$.
We have that $B_F$ \emph{simulates} $B_{F'}$ iff there exists a simulation relation $R$ such that $s_0'Rs_0$.
We pick $R = \corr^{-1}$.
Lemma \ref{lem:corrsim} already shows that $R$ is a simulation relation.
That $s_0'Rs_0$ follows easily from Lemma \ref{def:statecorr}, from which we can conclude that $B_F$ simulates $B_{F'}$.
\qed
\end{proof}

\noindent\textbf{Theorem \ref{the:safetymerge}.}
\textit{
Let $F$ and $F'$ be FIFO systems such that $F \cgtc F'$.
Then action-based safety properties are merge-preserved.
}
\begin{proof}
Follows from Theorem \ref{the:simulation} and Theorem 2 in \cite{DBLP:journals/fmsd/LoiseauxGSBB95}.
\qed
\end{proof}

\noindent\textbf{Lemma \ref{lem:reachsplit}.}
\textit{
Let $F = \langle P, C, M \rangle$ and $F' = \langle P, C', M \rangle$ be FIFO systems such that $F \cgtc F'$.
Let $B_F = \langle S, s_0, L, \trans{} \rangle$ and $B_{F'} = \langle S', s_0', L, \ptrans{} \rangle$.
Let $s \in S$ and $s' \in S'$.
Assume that $s \corr s'$.
Then $s \in Reach(S) \Leftarrow s' \in Reach(S')$.
}
\begin{proof}
This follows from Lemma \ref{lem:corrsim} and Theorem \ref{the:simulation}: if every sequence of actions from the initial state in $B_{F'}$ can be simulated by $B_F$ by following generalising states, then for every state reached this way, its generalising state in $B_F$ must be reachable as well along the same sequence of actions.
\qed
\end{proof}

\noindent\textbf{Lemma \ref{lem:reachmerge}.}
\textit{
Let $F = \langle P, C, M \rangle$ and $F' = \langle P, C', M \rangle$ be FIFO systems such that $F \cgtc F'$.
Let $B_F = \langle S, s_0, L, \trans{} \rangle$ and $B_{F'} = \langle S', s_0', L, \ptrans{} \rangle$.
Let $s \in S$ and $s' \in S'$.
Assume that $s \corr s'$.
Then $s \in Reach_{\IL(F, F')}(S) \Rightarrow s' \in Reach_{\IL(F, F')}(S')$.
}
\begin{proof}
We prove this by induction over the distance (in terms of minimal number of transitions) from the initial state.
As base case we have the initial state $s_0'$, which is trivially reachable by definition of reachability (Definition \ref{def:reachable}).
Now we prove the inductive step.
We assume that $s \in Reach_{IL(F, F')}(S)$ and is at distance $\delta$ from the initial state.
This means that there must be some state $t \in Reach_{IL(F, F')}(S)$ at distance $\delta - 1$ such that $t \trans{a} s$ for some $a \in L'$.
By the induction hypothesis, it follows that there exists a $t' \in Reach_{IL(F, F')}(S')$ for any $t' \in S'$ such that $t \corr t'$.
Then using Lemma \ref{lem:transcorrtau}, \ref{lem:transcorrrecnomerge} and \ref{lem:transcorrsend}, it follows for each $t'$ that $t' \ptrans{a} s'$ for some state $s' \in S'$ such that $s \corr s'$.
Since $a$ cannot change the contents of a merged channel, we know that the number of states that specialise $t$ equals the number of states that specialise $s$.
Combining the last two points, we know that for every $s'$ such that $s \corr s'$ there is a $t'$ such that $t' \in Reach_{IL(F, F')}(S')$ and $t' \ptrans{a} s'$, and therefore we can conclude that $s \in Reach_{IL(F, F')}(S) \Rightarrow s' \in Reach_{IL(F, F')}(S')$.
\qed
\end{proof}

\noindent\textbf{Theorem \ref{the:reachsplit}.}
\textit{
Let $F = \langle P, C, M \rangle$ and $F' = \langle P, C', M \rangle$ be FIFO systems such that $F \cgtc F'$.
Then for all $\kappa \in \mathbf{P}$, reachability of $\kappa$ is split-preserved.
}
\begin{proof}
Follows from Lemma \ref{lem:reachsplit} and \ref{lem:statecorr}.
\qed
\end{proof}

\noindent\textbf{Theorem \ref{the:reachmerge}.}
\textit{
Let $F = \langle P, C, M \rangle$ and $F' = \langle P, C', M \rangle$ be FIFO systems such that $F \cgtc F'$.
Then for all $\kappa \in \mathbf{P}$, $IL(F, F')$-reachability of $\kappa$ is merge-preserved.
}
\begin{proof}
Follows from Lemma \ref{lem:reachmerge} and \ref{lem:statecorr}.
\qed
\end{proof}

\noindent\textbf{Lemma \ref{lem:deadlocksplit}.}
\textit{
Let $F = \langle P, C, M \rangle$ and $F' = \langle P, C', M \rangle$ be FIFO systems such that $F \cgtc F'$.
Let $B_F = \langle S, s_0, L, \trans{} \rangle$ and $B_{F'} = \langle S', s_0', L, \ptrans{} \rangle$.
Let $s \in S$ and $s ' \in S'$.
Assume that $s \corr s'$.
Then $\neg\delta(s) \Leftarrow \neg\delta(s')$.
}
\begin{proof}
Follows from Lemma \ref{lem:corrsim}: if a transition is possible from $s'$, then $s$ is able to simulate it with a transition itself.
\qed
\end{proof}

\noindent\textbf{Theorem \ref{the:deadlocksplit}.}
\textit{
Let $F$ and $F'$ be FIFO systems such that $F \cgtc F'$.
Let $B_F = \langle S, s_0, L, \trans{} \rangle$ and $B_{F'} = \langle S', s_0', L, \ptrans{} \rangle$.
If $S \unre S'$, then deadlock freedom is split-preserved.
}
\begin{proof}
Follows from Lemma \ref{lem:deadlocksplit} and Theorem \ref{the:reachsplit}.
\qed
\end{proof}

\noindent\textbf{Lemma \ref{lem:conftau}.}
\textit{
Let $F = \langle P, C, M \rangle$ and $F' = \langle P, C', M \rangle$ be FIFO systems such that $F \cgtc F'$.
Let $B_F = \langle S, s_0, L, \trans{} \rangle$ and $B_{F'} = \langle S', s_0', L, \ptrans{} \rangle$.
Let $s \in S$, $s' \in S'$ and $a \in L$.
Assume that $s \corr s'$.
Then $\Conf^\tau_a(s) \Rightarrow \Conf^\tau_a(s')$ and if $\SC{a}$, then $\Conf^\tau_a(s) \Leftarrow \Conf^\tau_a(s')$.
}
\begin{proof}
Let $s = (\kappa_s, \zeta)$ and $s' = (\kappa_s, \zeta')$ for some $\kappa_s \in \mathbf{P}$, $\zeta \in \mathbf{C}$ and $\zeta' \in \mathbf{C}_\zeta'$.
We first prove $\Conf^\tau_a(s) \Rightarrow \Conf^\tau_a(s')$.
Assume that $\Conf^\tau_a(s)$, that is for all $t, u \in S$ we have that $(s \trans{\tau} t \wedge s \trans{a} u) \Rightarrow (\exists_{v \in S} : t \trans{a} v \wedge u \trans{\tau} v)$.
We need to prove that $\Conf^\tau_a(s')$, that is for all $t', u' \in S'$ we have that $(s' \ptrans{\tau} t' \wedge s' \ptrans{a} u') \Rightarrow (\exists_{v' \in S'} : t' \ptrans{a} v' \wedge u' \ptrans{\tau} v')$.
We do a case distinction on $a$.
\begin{itemize}
\item
Case $a = \tau$.
Pick some $t', u' \in S'$ such that $s' \ptrans{\tau} t'$ and $s' \ptrans{\tau} u'$.
Then by Definition \ref{def:FIFO2LTS}, $t' = (\kappa_t, \zeta')$ and $u' = (\kappa_u, \zeta')$ for some $\kappa_t, \kappa_u \in \mathbf{P}$.
Then using Lemma \ref{lem:transcorrtau}, we know that there are states $t, u \in S$ such that $t = (\kappa_t, \zeta)$, $u = (\kappa_u, \zeta)$, $s \trans{\tau} t$ and $s \trans{\tau} u$.
Due to $\Conf^\tau_a(s)$, we know that there must exist a $v \in S$, namely $v = (\kappa_v, \zeta)$ for some $\kappa_v \in \mathbf{P}$, such that $t \trans{\tau} v$ and $u \trans{\tau} v$.
Then using Lemma \ref{lem:transcorrtau}, we know that there must exist a $v' \in S'$, namely $v' = (\kappa_v, \zeta')$, such that $t' \ptrans{\tau} v'$ and $u' \ptrans{\tau} v'$, from which we can conclude that $\Conf^\tau_a(s')$.
\item
Case $a =\; ?m$ for some $m \in M$.
Pick some $t', u' \in S'$ such that $s' \ptrans{\tau} t'$ and $s' \ptrans{?m} u'$.
Then by Definition \ref{def:FIFO2LTS}, $hd(\zeta'(\chan{C'}{m})) = m$, $t' = (\kappa_t, \zeta')$ and $u' = (\kappa_u, \zeta'[\chan{C'}{m} \mapsto tl(\zeta'(\chan{C'}{m}))])$ for some $\kappa_t, \kappa_u \in \mathbf{P}$.
Then using Lemma \ref{lem:transcorrtau}-\ref{lem:transcorrrecmerge}, we know that there are states $t, u \in S$ such that $t = (\kappa_t, \zeta)$, $u = (\kappa_u, \zeta[\chan{C}{m} \mapsto tl(\zeta(\chan{C}{m}))])$, $s \trans{\tau} t$ and $s \trans{?m} u$.
Due to $\Conf^\tau_a(s)$, we know that there must exist a $v \in S$, namely $v = (\kappa_v, \zeta[\chan{C}{m} \mapsto tl(\zeta(\chan{C}{m}))])$ for some $\kappa_v \in \mathbf{P}$, such that $t \trans{?m} v$ and $u \trans{\tau} v$.
Then using Lemma \ref{lem:transcorrtau}-\ref{lem:transcorrrecmerge} and $hd(\zeta'(\chan{C'}{m})) = m$, we know that there must exist a $v' \in S'$, namely $v' = (\kappa_v, \zeta'[\chan{C'}{m} \mapsto tl(\zeta'(\chan{C'}{m}))])$, such that $t' \ptrans{?m} v'$ and $u' \ptrans{\tau} v'$, from which we can conclude that $\Conf^\tau_a(s')$.
\item
Case $a =\; !m$ for some $m \in M$.
Pick some $t', u' \in S'$ such that $s' \ptrans{\tau} t'$ and $s' \ptrans{!m} u'$.
Then by Definition \ref{def:FIFO2LTS}, $t' = (\kappa_t, \zeta')$ and $u' = (\kappa_u, \zeta'[\chan{C'}{m} \mapsto \zeta'(\chan{C'}{m}) \concat m])$ for some $\kappa_t, \kappa_u \in \mathbf{P}$.
Then using Lemma \ref{lem:transcorrtau} and \ref{lem:transcorrsend}, we know that there are states $t, u \in S$ such that $t = (\kappa_t, \zeta)$, $u = (\kappa_u, \zeta[\chan{C}{m} \mapsto \zeta(\chan{C}{m}) \concat m])$, $s \trans{\tau} t$ and $s \trans{!m} u$.
Due to $\Conf^\tau_a(s)$, we know that there must exist a $v \in S$, namely $v = (\kappa_v, \zeta[\chan{C}{m} \mapsto \zeta(\chan{C}{m}) \concat m])$ for some $\kappa_v \in \mathbf{P}$, such that $t \trans{!m} v$ and $u \trans{\tau} v$.
Then using Lemma \ref{lem:transcorrtau} and \ref{lem:transcorrsend}, we know that there must exist a $v' \in S'$, namely $v' = (\kappa_v, \zeta'[\chan{C'}{m} \mapsto \zeta'(\chan{C'}{m}) \concat m])$, such that $t' \ptrans{!m} v'$ and $u' \ptrans{\tau} v'$, from which we can conclude that $\Conf^\tau_a(s')$.
\end{itemize}

Next we prove that if $\SC{a}$, then $\Conf^\tau_a(s) \Leftarrow \Conf^\tau_a(s')$.
Assume that $\SC{a}$ and that $\Conf^\tau_a(s')$, that is for all $t', u' \in S'$ we have that $(s' \ptrans{\tau} t' \wedge s' \ptrans{a} u') \Rightarrow (\exists_{v' \in S'} : t' \ptrans{a} v' \wedge u' \ptrans{\tau} v')$.
We need to prove that $\Conf^\tau_a(s)$, that is for all $t, u \in S$ we have that $(s \trans{\tau} t \wedge s \trans{a} u) \Rightarrow (\exists_{v \in S} : t \trans{a} v \wedge u \trans{\tau} v)$.
We do a case distinction on $a$.
\begin{itemize}
\item
Case $a = \tau$.
Pick some $t, u \in S$ such that $s \trans{\tau} t$ and $s \trans{\tau} u$.
Then by Definition \ref{def:FIFO2LTS}, $t = (\kappa_t, \zeta)$ and $u = (\kappa_u, \zeta)$ for some $\kappa_t, \kappa_u \in \mathbf{P}$.
Then using Lemma \ref{lem:transcorrtau}, we know that there are states $t', u' \in S'$ such that $t' = (\kappa_t, \zeta')$, $u' = (\kappa_u, \zeta')$, $s' \ptrans{\tau} t'$ and $s' \ptrans{\tau} u'$.
Due to $\Conf^\tau_a(s')$, we know that there must exist a $v' \in S$, namely $v = (\kappa_v, \zeta')$ for some $\kappa_v \in \mathbf{P}$, such that $t' \ptrans{\tau} v'$ and $u' \ptrans{\tau} v'$.
Then using Lemma \ref{lem:transcorrtau}, we know that there must exist a $v \in S$, namely $v = (\kappa_v, \zeta)$, such that $t \trans{\tau} v$ and $u \trans{\tau} v$, from which we can conclude that $\Conf^\tau_a(s)$.
\item
Case $a =\; ?m$ for some $m \in M$.
Pick some $t, u \in S$ such that $s \trans{\tau} t$ and $s \trans{?m} u$.
Then by Definition \ref{def:FIFO2LTS}, $t = (\kappa_t, \zeta)$ and $u = (\kappa_u, \zeta[\chan{C}{m} \mapsto tl(\zeta(\chan{C}{m}))])$ for some $\kappa_t, \kappa_u \in \mathbf{P}$.
Then using $\SC{a}$ and Lemma \ref{lem:transcorrtau} and \ref{lem:transcorrrecnomerge}, we know that there are states $t', u' \in S'$ such that $t' = (\kappa_t, \zeta')$, $u' = (\kappa_u, \zeta'[\chan{C'}{m} \mapsto tl(\zeta'(\chan{C'}{m}))])$, $s' \ptrans{\tau} t'$ and $s' \ptrans{?m} u'$.
Due to $\Conf^\tau_a(s')$, we know that there must exist a $v' \in S'$, namely $v' = (\kappa_v, \zeta'[\chan{C'}{m} \mapsto tl(\zeta'(\chan{C'}{m}))])$ for some $\kappa_v \in \mathbf{P}$, such that $t' \ptrans{?m} v'$ and $u' \ptrans{\tau} v'$.
Then using $\SC{a}$ and Lemma \ref{lem:transcorrtau} and \ref{lem:transcorrrecnomerge}, we know that there must exist a $v \in S$, namely $v = (\kappa_v, \zeta[\chan{C}{m} \mapsto tl(\zeta(\chan{C}{m}))])$, such that $t \trans{m} v$ and $u \trans{\tau} v$, from which we can conclude that $\Conf^\tau_a(s)$.
\item
Case $a =\; !m$ for some $m \in M$.
Pick some $t, u \in S$ such that $s \trans{\tau} t$ and $s \trans{!m} u$.
Then by Definition \ref{def:FIFO2LTS}, $t = (\kappa_t, \zeta)$ and $u = (\kappa_u, \zeta[\chan{C}{m} \mapsto \zeta(\chan{C}{m}) \concat m])$ for some $\kappa_t, \kappa_u \in \mathbf{P}$.
Then using Lemma \ref{lem:transcorrtau} and \ref{lem:transcorrsend}, we that there are states $t', u' \in S'$ such that $t' = (\kappa_t, \zeta')$, $u' = (\kappa_u, \zeta'[\chan{C'}{m} \mapsto \zeta'(\chan{C'}{m}) \concat m])$, $s' \ptrans{\tau} t'$ and $s' \ptrans{!m} u'$.
Due to $\Conf^\tau_a(s')$, we know that there must exist a $v' \in S'$, namely $v' = (\kappa_v, \zeta'[\chan{C'}{m} \mapsto \zeta'(\chan{C'}{m}) \concat m])$ for some $\kappa_v \in \mathbf{P}$, such that $t' \ptrans{!m} v'$ and $u' \ptrans{\tau} v'$.
Then using Lemma \ref{lem:transcorrtau} and \ref{lem:transcorrsend}, we know that there must exist a $v \in S$, namely $v = (\kappa_v, \zeta[\chan{C}{m} \mapsto \zeta(\chan{C}{m}) \concat m])$, such that $t \trans{!m} v$ and $u \trans{\tau} v$, from which we can conclude that $\Conf^\tau_a(s)$.
\end{itemize}
\qed
\end{proof}

\noindent\textbf{Lemma \ref{lem:confoc}.}
\textit{
Let $F = \langle P, C, M \rangle$ and $F' = \langle P, C', M \rangle$ be FIFO systems such that $F \cgtc F'$.
Let $B_F = \langle S, s_0, L, \trans{} \rangle$ and $B_{F'} = \langle S', s_0', L, \ptrans{} \rangle$.
Let $s \in S$, $s' \in S'$ and $m, o \in M$.
Assume that $s \corr s'$ and $m \not\simeq_{C'} o$.
Let $a \in \{?m, !m\}$ and $b \in \{?o, !o\}$.
Then $\Conf^a_b(s) \Rightarrow \Conf^a_b(s')$ and if $\SC{a}$ and $\SC{b}$, then $\Conf^a_b(s) \Leftarrow \Conf^a_b(s')$.
}
\begin{proof}
Note that $m \not\simeq_{C'} o$ implies that $m \not\simeq_C o$.
Let $s = (\kappa_s, \zeta)$ and $s' = (\kappa_s, \zeta')$ for some $\kappa_s \in \mathbf{P}$, $\zeta \in \mathbf{C}$ and $\zeta' \in \mathbf{C}_\zeta'$.
We first prove $\Conf^a_b(s) \Rightarrow \Conf^a_b(s')$.
Assume that $\Conf^a_b(s)$, that is for all $t, u \in S$ we have that $(s \trans{a} t \wedge s \trans{b} u) \Rightarrow (\exists_{v \in S} : t \trans{b} v \wedge u \trans{a} v)$.
We need to prove that $\Conf^a_b(s')$, that is for all $t', u' \in S'$ we have that $(s' \ptrans{a} t' \wedge s' \ptrans{b} u') \Rightarrow (\exists_{v' \in S'} : t' \ptrans{b} v' \wedge u' \ptrans{a} v')$.
We do a case distinction on $a$ and $b$.
\begin{itemize}
\item
Case $a =\; ?m$ and $b =\; ?o$.
Pick some $t', u' \in S'$ such that $s' \ptrans{?m} t'$ and $s' \ptrans{?o} u'$.
Then by Definition \ref{def:FIFO2LTS}, $hd(\zeta'(\chan{C'}{m})) = m$, $hd(\zeta'(\chan{C'}{o})) = o$, $t' = (\kappa_t, \zeta'[\chan{C'}{m} \mapsto tl(\zeta'(\chan{C'}{m}))])$ and $u' = (\kappa_u, \zeta'[\chan{C'}{o} \mapsto tl(\zeta'(\chan{C'}{o}))])$ for some $\kappa_t, \kappa_u \in \mathbf{P}$.
Then using Lemma \ref{lem:transcorrrecnomerge} and \ref{lem:transcorrrecmerge}, we know that there are states $t, u \in S$ such that $t = (\kappa_t, \zeta[\chan{C}{m} \mapsto tl(\zeta(\chan{C}{m}))])$, $u = (\kappa_u, \zeta[\chan{C}{o} \mapsto tl(\zeta(\chan{C}{o}))])$, $s \trans{?m} t$ and $s \trans{?o} u$.
Due to $\Conf^a_b(s)$ and $m \not\simeq_C o$, we know that there must exist a $v \in S$, namely $v = (\kappa_v, \zeta[\chan{C}{m} \mapsto tl(\zeta(\chan{C}{m})), \chan{C}{o} \mapsto tl(\zeta(\chan{C}{o}))])$ for some $\kappa_v \in \mathbf{P}$, such that $t \trans{?o} v$ and $u \trans{?m} v$.
Then using $hd(\zeta'(\chan{C'}{m})) = m$, $hd(\zeta'(\chan{C'}{o})) = o$, $m \not\simeq_{C'} o$ and Lemma \ref{lem:transcorrrecnomerge} and \ref{lem:transcorrrecmerge}, we know that there must exist a $v' \in S'$, namely $v' = (\kappa_v, \zeta'[\chan{C'}{m} \mapsto tl(\zeta'(\chan{C'}{m})), \chan{C'}{o} \mapsto tl(\zeta'(\chan{C'}{o}))])$, such that $t' \ptrans{?o} v'$ and $u' \ptrans{?m} v'$, from which we can conclude that $\Conf^a_b(s')$.
\item
Case $a =\; !m$ and $b =\; !o$.
Pick some $t', u' \in S'$ such that $s' \ptrans{!m} t'$ and $s' \ptrans{!o} u'$.
Then by Definition \ref{def:FIFO2LTS}, $t' = (\kappa_t, \zeta'[\chan{C'}{m} \mapsto \zeta'(\chan{C'}{m}) \concat m])$ and $u' = (\kappa_u, \zeta'[\chan{C'}{o} \mapsto \zeta'(\chan{C'}{o}) \concat o])$ for some $\kappa_t, \kappa_u \in \mathbf{P}$.
Then using Lemma \ref{lem:transcorrsend}, we know there are states $t, u \in S$ such that $t = (\kappa_t, \zeta[\chan{C}{m} \mapsto \zeta(\chan{C}{m}) \concat m])$, $u = (\kappa_u, \zeta[\chan{C}{o} \mapsto \zeta(\chan{C}{o}) \concat o])$, $s \trans{!m} t$ and $s \trans{!o} u$.
Due to $\Conf^a_b(s)$ and $m \not\simeq_C o$, we know that there must exist a $v \in S$, namely $v = (\kappa_v, \zeta[\chan{C}{m} \mapsto \zeta(\chan{C}{m}) \concat m, \chan{C}{o} \mapsto \zeta(\chan{C}{o}) \concat o])$ for some $\kappa_v \in \mathbf{P}$, such that $t \trans{!o} v$ and $u \trans{!m} v$.
Then using $m \not\simeq_{C'} o$ and Lemma \ref{lem:transcorrsend}, we know that there must exist a $v' \in S'$, namely $v' = (\kappa_v, \zeta'[\chan{C'}{m} \mapsto \zeta'(\chan{C'}{m}) \concat m, \chan{C'}{o} \mapsto \zeta'(\chan{C'}{o}) \concat o])$, such that $t' \ptrans{!o} v'$ and $u' \ptrans{!m} v'$, from which we can conclude that $\Conf^a_b(s')$.
\item
Case $a =\; ?m$ and $b =\; !o$.
Pick some $t', u' \in S'$ such that $s' \ptrans{?m} t'$ and $s' \ptrans{!o} u'$.
Then by Definition \ref{def:FIFO2LTS}, $hd(\zeta'(\chan{C'}{m})) = m$, $t' = (\kappa_t, \zeta'[\chan{C'}{m} \mapsto tl(\zeta'(\chan{C'}{m}))])$ and $u' = (\kappa_u, \zeta'[\chan{C'}{o} \mapsto \zeta'(\chan{C'}{o}) \concat o])$ for some $\kappa_t, \kappa_u \in \mathbf{P}$.
Then using Lemma \ref{lem:transcorrrecnomerge}-\ref{lem:transcorrsend}, we know that there are states $t, u \in S$ such that $t = (\kappa_t, \zeta[\chan{C}{m} \mapsto tl(\zeta(\chan{C}{m}))])$, $u = (\kappa_u, \zeta[\chan{C}{o} \mapsto \zeta(\chan{C}{o}) \concat o])$, $s \trans{?m} t$ and $s \trans{!o} u$.
Due to $\Conf^a_b(s)$ and $m \not\simeq_C o$, we know that there must exist a $v \in S$, namely $v = (\kappa_v, \zeta[\chan{C}{m} \mapsto tl(\zeta(\chan{C}{m})), \chan{C}{o} \mapsto \zeta(\chan{C}{o}) \concat o])$ for some $\kappa_v \in \mathbf{P}$, such that $t \trans{!o} v$ and $u \trans{!m} v$.
Then using $hd(\zeta'(\chan{C'}{m})) = m$, $m \not\simeq_{C'} o$ and Lemma \ref{lem:transcorrrecnomerge}-\ref{lem:transcorrsend}, we know that there must exist a $v' \in S'$, namely $v' = (\kappa_v, \zeta'[\chan{C'}{m} \mapsto tl(\zeta'(\chan{C'}{m})), \chan{C'}{o} \mapsto \zeta'(\chan{C'}{o}) \concat o])$ such that $t' \ptrans{!o} v'$ and $u' \ptrans{!m} v'$, from which we can conclude that $\Conf^a_b(s')$.
\item
Case $a =\; !m$ and $b =\; ?o$.
This case follows by symmetry of the above case.
\end{itemize}

Next we prove that if $\SC{a}$ and $\SC{b}$, then $\Conf^a_b(s) \Leftarrow \Conf^a_b(s')$.
Assume that $\SC{a}$, $\SC{b}$ and that $\Conf^a_b(s')$, that is for all $t', u' \in S'$ we have that $(s' \ptrans{a} t' \wedge s' \ptrans{b} u') \Rightarrow (\exists_{v' \in S'} : t' \ptrans{b} v' \wedge u' \ptrans{a} v')$.
We need to prove that $\Conf^a_b(s)$, that is for all $t, u \in S$ we have that $(s \trans{a} t \wedge s \trans{b} u) \Rightarrow (\exists_{v \in S} : t \trans{b} v \wedge u \trans{a} v)$.
We do a case distinction on $a$ and $b$.
\begin{itemize}
\item
Case $a =\; ?m$ and $b =\; ?o$.
Pick some $t, u \in S$ such that $s \trans{?m} t$ and $s \trans{?o} u$.
Then by Definition \ref{def:FIFO2LTS}, $t = (\kappa_t, \zeta[\chan{C}{m} \mapsto tl(\zeta(\chan{C}{m}))])$ and $u = (\kappa_u, \zeta[\chan{C'}{o} \mapsto tl(\zeta(\chan{C}{o}))])$ for some $\kappa_t, \kappa_u \in \mathbf{P}$.
Then using $\SC{a}$, $\SC{b}$ and Lemma \ref{lem:transcorrrecnomerge}, we know that there are states $t', u' \in S'$ such that $t' = (\kappa_t, \zeta'[\chan{C'}{m} \mapsto tl(\zeta'(\chan{C'}{m}))])$, $u' = (\kappa_u, \zeta'[\chan{C'}{o} \mapsto tl(\zeta'(\chan{C'}{o}))])$, $s' \ptrans{?m} t'$ and $s' \ptrans{?o} u'$.
Due to $\Conf^a_b(s')$ and $m \not\simeq_{C'} o$, we know that there must exist a $v' \in S'$, namely $v' = (\kappa_v, \zeta'[\chan{C'}{m} \mapsto tl(\zeta'(\chan{C'}{m})), \chan{C'}{o} \mapsto tl(\zeta'(\chan{C'}{o}))])$ for some $\kappa_v \in \mathbf{P}$, such that $t' \ptrans{?o} v'$ and $u' \ptrans{?m} v'$.
Then using $\SC{a}$, $\SC{b}$, $m \not\simeq_C o$ and Lemma \ref{lem:transcorrrecnomerge} and \ref{lem:transcorrrecmerge}, we know that there must exist a $v \in S$, namely $v = (\kappa_v, \zeta[\chan{C}{m} \mapsto tl(\zeta(\chan{C}{m})), \chan{C}{o} \mapsto tl(\zeta(\chan{C}{o}))])$, such that $t \trans{?o} v$ and $u \trans{?m} v$, from which we can conclude that $\Conf^a_b(s)$.
\item
Case $a =\; !m$ and $b =\; !o$.
Pick some $t, u \in S$ such that $s \trans{!m} t$ and $s \trans{!o} u$.
Then by Definition \ref{def:FIFO2LTS}, $t = (\kappa_t, \zeta[\chan{C}{m} \mapsto \zeta(\chan{C}{m}) \concat m])$ and $u = (\kappa_u, \zeta[\chan{C}{o} \mapsto \zeta(\chan{C}{o}) \concat o])$ for some $\kappa_t, \kappa_u \in \mathbf{P}$.
Then using Lemma \ref{lem:transcorrsend}, we know there are states $t', u' \in S'$ such that $t' = (\kappa_t, \zeta'[\chan{C'}{m} \mapsto \zeta'(\chan{C'}{m}) \concat m])$, $u' = (\kappa_u, \zeta'[\chan{C'}{o} \mapsto \zeta'(\chan{C'}{o}) \concat o])$, $s' \ptrans{!m} t'$ and $s' \ptrans{!o} u'$.
Due to $\Conf^a_b(s')$ and $m \not\simeq_{C'} o$, we know that there must exist a $v' \in S'$, namely $v' = (\kappa_v, \zeta'[\chan{C'}{m} \mapsto \zeta'(\chan{C'}{m}) \concat m, \chan{C'}{o} \mapsto \zeta'(\chan{C'}{o}) \concat o])$ for some $\kappa_v \in \mathbf{P}$, such that $t' \ptrans{!o} v'$ and $u' \ptrans{!m} v'$.
Then using $m \not\simeq_C o$ and Lemma \ref{lem:transcorrsend}, we know that there must exist a $v \in S$, namely $v = (\kappa_v, \zeta[\chan{C}{m} \mapsto \zeta(\chan{C}{m}) \concat m, \chan{C}{o} \mapsto \zeta(\chan{C}{o}) \concat o])$, such that $t \trans{!o} v$ and $u \trans{!m} v$, from which we can conclude that $\Conf^a_b(s)$.
\item
Case $a =\; ?m$ and $b =\; !o$.
Pick some $t, u \in S$ such that $s \trans{?m} t$ and $s \trans{!o} u$.
Then by Definition \ref{def:FIFO2LTS}, $t = (\kappa_t, \zeta[\chan{C}{m} \mapsto tl(\zeta(\chan{C}{m}))])$ and $u = (\kappa_u, \zeta[\chan{C}{o} \mapsto \zeta(\chan{C}{o}) \concat o])$ for some $\kappa_t, \kappa_u \in \mathbf{P}$.
Then using $\SC{a}$ and Lemma \ref{lem:transcorrrecnomerge} and \ref{lem:transcorrsend}, we know that there are states $t', u' \in S'$ such that $t' = (\kappa_t, \zeta'[\chan{C'}{m} \mapsto tl(\zeta'(\chan{C'}{m}))])$, $u' = (\kappa_u, \zeta'[\chan{C'}{o} \mapsto \zeta'(\chan{C'}{o}) \concat o])$, $s' \ptrans{?m} t'$ and $s' \ptrans{!o} u'$.
Due to $\Conf^a_b(s')$ and $m \not\simeq_{C'} o$, we know that there must exist a $v' \in S'$, namely $v' = (\kappa_v, \zeta'[\chan{C'}{m} \mapsto tl(\zeta'(\chan{C'}{m})), \chan{C'}{o} \mapsto \zeta'(\chan{C'}{o}) \concat o])$ for some $\kappa_v \in \mathbf{P}$, such that $t' \ptrans{!o} v'$ and $u' \ptrans{!m} v'$.
Then using $\SC{a}$, $m \not\simeq_C o$ and Lemma \ref{lem:transcorrrecnomerge} and \ref{lem:transcorrsend}, we know that there must exist a $v \in S$, namely $v = (\kappa_v, \zeta[\chan{C}{m} \mapsto tl(\zeta(\chan{C}{m})), \chan{C}{o} \mapsto \zeta(\chan{C}{o}) \concat o])$ such that $t \trans{!o} v$ and $u \trans{!m} v$, from which we can conclude that $\Conf^a_b(s)$.
\item
Case $a =\; !m$ and $b =\; ?o$.
This case follows by symmetry of the above case.
\end{itemize}
\qed
\end{proof}

\noindent\textbf{Lemma \ref{lem:confsc}.}
\textit{
Let $F = \langle P, C, M \rangle$ and $F' = \langle P, C', M \rangle$ be FIFO systems such that $F \cgtc F'$.
Let $B_F = \langle S, s_0, L, \trans{} \rangle$ and $B_{F'} = \langle S', s_0', L, \ptrans{} \rangle$.
Let $s \in S$, $s' \in S'$ and $m, o \in M$.
Assume that $s \corr s'$ and $m \simeq_C o$.
Let $a \in \{?m, !m\}$ and $b \in \{?o, !o\}$.
If not $a \equiv_? b$, then $\Conf^a_b(s) \Rightarrow \Conf^a_b(s')$ and if $\SC{a}$ and $\SC{b}$, then $\Conf^a_b(s) \Leftarrow \Conf^a_b(s')$.
}
\begin{proof}
Note that $m \simeq_C o$ implies that $m \simeq_{C'} o$.
Let $s = (\kappa_s, \zeta)$ and $s' = (\kappa_s, \zeta')$ for some $\kappa_s \in \mathbf{P}$, $\zeta \in \mathbf{C}$ and $\zeta' \in \mathbf{C}_\zeta'$.
We first prove that if not $a \equiv_? b$, then $\Conf^a_b(s) \Rightarrow \Conf^a_b(s')$.
Assume that not $a \equiv_? b$ and that $\Conf^a_b(s)$, that is for all $t, u \in S$ we have that $(s \trans{a} t \wedge s \trans{b} u) \Rightarrow (\exists_{v \in S} : t \trans{b} v \wedge u \trans{a} v)$.
We need to prove that $\Conf^a_b(s')$, that is for all $t', u' \in S'$ we have that $(s' \ptrans{a} t' \wedge s' \ptrans{b} u') \Rightarrow (\exists_{v' \in S'} : t' \ptrans{b} v' \wedge u' \ptrans{a} v')$.
We do a case distinction on $a$ and $b$.
\begin{itemize}
\item
Case $a =\; ?m$ and $b =\; ?o$.
Since not $a \equiv_? b$, we know that $m \neq o$.
Due to $m \neq o$ and $m \simeq_{C'} o$, the choice between actions $?m$ and $?o$ is not possible from $s'$, because a channel cannot have two distinct messages at its head.
Therefore $\Conf^a_b(s')$ trivially holds.
\item
Case $a =\; !m$ and $b =\; !o$.
In case $m \neq o$, since $m \simeq_C o$, there does not exist a fourth state to complete the confluence diamond from $s$, because different orderings of $!m$ and $!o$ will result in different channel orderings.
Therefore $\Conf^a_b(s)$ cannot hold, which makes the implication hold trivially.

In case $m = o$, pick some $t', u' \in S'$ such that $s' \ptrans{!m} t'$ and $s' \ptrans{!m} u'$.
Then by Definition \ref{def:FIFO2LTS}, $t' = (\kappa_t, \zeta'[\chan{C'}{m} \mapsto \zeta'(\chan{C'}{m}) \concat m])$ and $u' = (\kappa_u, \zeta'[\chan{C'}{m} \mapsto \zeta'(\chan{C'}{m}) \concat m])$ for some $\kappa_t, \kappa_u \in \mathbf{P}$.
Then using Lemma \ref{lem:transcorrsend}, we know there are states $t, u \in S$ such that $t = (\kappa_t, \zeta[\chan{C}{m} \mapsto \zeta(\chan{C}{m}) \concat m])$, $u = (\kappa_u, \zeta[\chan{C}{m} \mapsto \zeta(\chan{C}{m}) \concat m])$, $s \trans{!m} t$ and $s \trans{!o} u$.
Due to $\Conf^a_b(s)$, we know that there must exist a $v \in S$, namely $v = (\kappa_v, \zeta[\chan{C}{m} \mapsto \zeta(\chan{C}{m}) \concat m \concat m])$ for some $\kappa_v \in \mathbf{P}$, such that $t \trans{!m} v$ and $u \trans{!m} v$.
Then using Lemma \ref{lem:transcorrsend}, we know that there must exist a $v' \in S'$, namely $v' = (\kappa_v, \zeta'[\chan{C'}{m} \mapsto \zeta'(\chan{C'}{m}) \concat m \concat m])$, such that $t' \ptrans{!m} v'$ and $u' \ptrans{!m} v'$, from which we can conclude that $\Conf^a_b(s')$.
\item
Case $a =\; ?m$ and $b =\; !o$.
Pick some $t', u' \in S'$ such that $s' \ptrans{?m} t'$ and $s' \ptrans{!o} u'$.
Then by Definition \ref{def:FIFO2LTS}, $hd(\zeta'(\chan{C'}{m})) = m$, $t' = (\kappa_t, \zeta'[\chan{C'}{m} \mapsto tl(\zeta'(\chan{C'}{m}))])$ and $u' = (\kappa_u, \zeta'[\chan{C'}{o} \mapsto \zeta'(\chan{C'}{o}) \concat o])$ for some $\kappa_t, \kappa_u \in \mathbf{P}$.
Then using Lemma \ref{lem:transcorrrecnomerge}-\ref{lem:transcorrsend}, we know that there are states $t, u \in S$ such that $t = (\kappa_t, \zeta[\chan{C}{m} \mapsto tl(\zeta(\chan{C}{m}))])$, $u = (\kappa_u, \zeta[\chan{C}{o} \mapsto \zeta(\chan{C}{o}) \concat o])$, $s \trans{?m} t$ and $s \trans{!o} u$.
Due to $\Conf^a_b(s)$, $m \simeq_C o$ and $tl(w) \concat m = tl(w \concat m)$ for any $m \in M$ and $w \in M^*$, we know that there must exist a $v \in S$, namely $v = (\kappa_v, \zeta[\chan{C}{m} \mapsto tl(\zeta(\chan{C}{m})) \concat o])$ for some $\kappa_v \in \mathbf{P}$, such that $t \trans{!o} v$ and $u \trans{!m} v$.
Then using $hd(\zeta'(\chan{C'}{m})) = m$, $m \simeq_{C'} o$, $tl(w) \concat m = tl(w \concat m)$ for any $m \in M$ and $w \in M^*$ and Lemma \ref{lem:transcorrrecnomerge}-\ref{lem:transcorrsend}, we know that there must exist a $v' \in S'$, namely $v' = (\kappa_v, \zeta'[\chan{C'}{m} \mapsto tl(\zeta'(\chan{C'}{m})) \concat o])$ such that $t' \ptrans{!o} v'$ and $u' \ptrans{!m} v'$, from which we can conclude that $\Conf^a_b(s')$.
\item
Case $a =\; !m$ and $b =\; ?o$.
This case follows by symmetry of the above case.
\end{itemize}

Next we prove that if $\SC{a}$ and $\SC{b}$, then $\Conf^a_b(s) \Leftarrow \Conf^a_b(s')$.
Assume that $\SC{a}$ and $\SC{b}$ and that $\Conf^a_b(s')$, that is for all $t', u' \in S'$ we have that $(s' \ptrans{a} t' \wedge s' \ptrans{b} u') \Rightarrow (\exists_{v' \in S'} : t' \ptrans{b} v' \wedge u' \ptrans{a} v')$.
We need to prove that $\Conf^a_b(s)$, that is for all $t, u \in S$ we have that $(s \trans{a} t \wedge s \trans{b} u) \Rightarrow (\exists_{v \in S} : t \trans{b} v \wedge u \trans{a} v)$.
We do a case distinction on $a$ and $b$.
\begin{itemize}
\item
Case $a =\; ?m$ and $b =\; ?o$.
Since not $a \equiv_? b$, we know that $m \neq o$.
Due to $m \neq o$ and $m \simeq_C o$, the choice between actions $?m$ and $?o$ is not possible from $s$, because a channel cannot have two distinct messages at its head.
Therefore $\Conf^a_b(s)$ trivially holds.
\item
Case $a =\; !m$ and $b =\; !o$.
In case $m \neq o$, since $m \simeq_{C'} o$, there does not exist a fourth state to complete the confluence diamond from $s'$, because different orderings of $!m$ and $!o$ will result in different channel orderings.
Therefore $\Conf^a_b(s')$ cannot hold, which makes the implication hold trivially.

In case $m = o$, pick some $t, u \in S$ such that $s \trans{!m} t$ and $s \trans{!m} u$.
Then by Definition \ref{def:FIFO2LTS}, $t = (\kappa_t, \zeta[\chan{C}{m} \mapsto \zeta(\chan{C}{m}) \concat m])$ and $u = (\kappa_u, \zeta[\chan{C}{m} \mapsto \zeta(\chan{C}{m}) \concat m])$ for some $\kappa_t, \kappa_u \in \mathbf{P}$.
Then using Lemma \ref{lem:transcorrsend}, we know there are states $t', u' \in S'$ such that $t' = (\kappa_t, \zeta'[\chan{C'}{m} \mapsto \zeta'(\chan{C'}{m}) \concat m])$, $u' = (\kappa_u, \zeta'[\chan{C'}{m} \mapsto \zeta'(\chan{C'}{m}) \concat m])$, $s' \ptrans{!m} t'$ and $s' \ptrans{!o} u'$.
Due to $\Conf^a_b(s')$, we know that there must exist a $v' \in S'$, namely $v' = (\kappa_v, \zeta'[\chan{C'}{m} \mapsto \zeta'(\chan{C'}{m}) \concat m \concat m])$ for some $\kappa_v \in \mathbf{P}$, such that $t' \ptrans{!m} v'$ and $u' \ptrans{!m} v'$.
Then using Lemma \ref{lem:transcorrsend}, we know that there must exist a $v \in S$, namely $v = (\kappa_v, \zeta[\chan{C}{m} \mapsto \zeta(\chan{C}{m}) \concat m \concat m])$, such that $t \trans{!m} v$ and $u \trans{!m} v$, from which we can conclude that $\Conf^a_b(s)$.
\item
Case $a =\; ?m$ and $b =\; !o$.
Pick some $t, u \in S$ such that $s \trans{?m} t$ and $s \trans{!o} u$.
Then by Definition \ref{def:FIFO2LTS}, $t = (\kappa_t, \zeta[\chan{C}{m} \mapsto tl(\zeta(\chan{C}{m}))])$ and $u = (\kappa_u, \zeta[\chan{C}{o} \mapsto \zeta(\chan{C}{o}) \concat o])$ for some $\kappa_t, \kappa_u \in \mathbf{P}$.
Then using $\SC{a}$ and Lemma \ref{lem:transcorrrecnomerge} and \ref{lem:transcorrsend}, we know that there are states $t', u' \in S'$ such that $t' = (\kappa_t, \zeta'[\chan{C'}{m} \mapsto tl(\zeta'(\chan{C'}{m}))])$, $u' = (\kappa_u, \zeta'[\chan{C'}{o} \mapsto \zeta'(\chan{C'}{o}) \concat o])$, $s' \ptrans{?m} t'$ and $s' \ptrans{!o} u'$.
Due to $\Conf^a_b(s')$, $m \simeq_{C'} o$ and $tl(w) \concat m = tl(w \concat m)$ for any $m \in M$ and $w \in M^*$, we know that there must exist a $v' \in S'$, namely $v' = (\kappa_v, \zeta'[\chan{C'}{m} \mapsto tl(\zeta'(\chan{C'}{m})) \concat o])$ for some $\kappa_v \in \mathbf{P}$, such that $t' \ptrans{!o} v'$ and $u' \ptrans{!m} v'$.
Then using $\SC{a}$, $m \simeq_C o$, $tl(w) \concat m = tl(w \concat m)$ for any $m \in M$ and $w \in M^*$ and Lemma \ref{lem:transcorrrecnomerge} and \ref{lem:transcorrsend}, we know that there must exist a $v \in S$, namely $v = (\kappa_v, \zeta[\chan{C}{m} \mapsto tl(\zeta(\chan{C}{m})) \concat o])$ such that $t \trans{!o} v$ and $u \trans{!m} v$, from which we can conclude that $\Conf^a_b(s)$.
\item
Case $a =\; !m$ and $b =\; ?o$.
This case follows by symmetry of the above case.
\end{itemize}
\qed
\end{proof}

\noindent\textbf{Lemma \ref{lem:confccmerge}.}
\textit{
Let $F = \langle P, C, M \rangle$ and $F' = \langle P, C', M \rangle$ be FIFO systems such that $F \cgtc F'$.
Let $B_F = \langle S, s_0, L, \trans{} \rangle$ and $B_{F'} = \langle S', s_0', L, \ptrans{} \rangle$.
Let $s \in S$, $s' \in S'$ and $m, o \in M$.
Assume that $s \corr s'$, $m \not\simeq_C o$ and $m \simeq_{C'} o$.
Let $a \in \{?m, !m\}$ and $b \in \{?o, !o\}$.
If not $\AC{a}{b}$, then $\Conf^{a}_{b}(s) \Rightarrow \Conf^{a}_{b}(s')$.
}
\begin{proof}
Let $s = (\kappa_s, \zeta)$ and $s' = (\kappa_s, \zeta')$ for some $\kappa_s \in \mathbf{P}$, $\zeta \in \mathbf{C}$ and $\zeta' \in \mathbf{C}_\zeta'$.
Assume that $\Conf^a_b(s)$, that is for all $t, u \in S$ we have that $(s \trans{a} t \wedge s \trans{b} u) \Rightarrow (\exists_{v \in S} : t \trans{b} v \wedge u \trans{a} v)$.
We need to prove that $\Conf^a_b(s')$, that is for all $t', u' \in S'$ we have that $(s' \ptrans{a} t' \wedge s' \ptrans{b} u') \Rightarrow (\exists_{v' \in S'} : t' \ptrans{b} v' \wedge u' \ptrans{a} v')$.
We do a case distinction on $a$ and $b$.
Note that the case where $a = !m$ and $b = !o$ is not possible since we assume that not $\AC{a}{b}$.
\begin{itemize}
\item
Case $a =\; ?m$ and $b =\; ?o$.
From $m \not\simeq_C o$ it follows that $m \neq o$.
Since $m \simeq_{C'} o$ and $m \neq o$ and since a channel can only have one element as its head, the choice between $?m$ and $?o$ is not possible from $s'$, so $\Conf^a_b(s')$ trivially holds.
\item
Case $a =\; ?m$ and $b =\; !o$.
Pick some $t', u' \in S'$ such that $s' \ptrans{?m} t'$ and $s' \ptrans{!o} u'$.
Then by Definition \ref{def:FIFO2LTS}, $hd(\zeta'(\chan{C'}{m})) = m$, $t' = (\kappa_t, \zeta'[\chan{C'}{m} \mapsto tl(\zeta'(\chan{C'}{m}))])$ and $u' = (\kappa_u, \zeta'[\chan{C'}{o} \mapsto \zeta'(\chan{C'}{o}) \concat o])$ for some $\kappa_t, \kappa_u \in \mathbf{P}$.
Then using Lemma \ref{lem:transcorrrecmerge} and \ref{lem:transcorrsend}, we know that there are states $t, u \in S$ such that $t = (\kappa_t, \zeta[\chan{C}{m} \mapsto tl(\zeta(\chan{C}{m}))])$, $u = (\kappa_u, \zeta[\chan{C}{o} \mapsto \zeta(\chan{C}{o}) \concat o])$, $s \trans{?m} t$ and $s \trans{!o} u$.
Due to $\Conf^{?m}_{!o}(s)$, $m \not\simeq_C o$ and $tl(w) \concat m = tl(w \concat m)$ for any $m \in M$ and $w \in M^*$, we know that there must exist a $v \in S$, namely $v = (\kappa_v, \zeta[\chan{C}{m} \mapsto tl(\zeta(\chan{C}{m})), \chan{C}{o} \mapsto \zeta(\chan{C}{o}) \concat o])$ for some $\kappa_v \in \mathbf{P}$, such that $t \trans{!o} v$ and $u \trans{?m} v$.
Then using $hd(\zeta'(\chan{C'}{m})) = m$, $m \simeq_{C'} o$, $tl(w) \concat m = tl(w \concat m)$ for any $m \in M$ and $w \in M^*$ and Lemma \ref{lem:transcorrrecmerge} and \ref{lem:transcorrsend}, we know that there must exist a $v' \in S'$, namely $v' = (\kappa_v, \zeta'[\chan{C'}{m} \mapsto tl(\zeta'(\chan{C'}{m})) \concat o])$ such that $t' \ptrans{!o} v'$ and $u' \ptrans{?m} v'$, from which we can conclude that $\Conf^{?m}_{!o}(s')$.
\item
Case $a =\; !m$ and $b =\; ?o$.
This case follows by symmetry of the above case.
\end{itemize}
\qed
\end{proof}

\noindent\textbf{Lemma \ref{lem:confccsplit}.}
\textit{
Let $F = \langle P, C, M \rangle$ and $F' = \langle P, C', M \rangle$ be FIFO systems such that $F \cgtc F'$.
Let $B_F = \langle S, s_0, L, \trans{} \rangle$ and $B_{F'} = \langle S', s_0', L, \ptrans{} \rangle$.
Let $s \in S$, $s' \in S'$ and $a, b \in L$
Assume that $s \corr s'$.
If $\AC{a}{b}$, then $\Conf^a_b(s) \Leftarrow \Conf^a_b(s')$.
}
\begin{proof}
Since $m \simeq_{C'} o$, there does not exist a fourth state to complete the confluence diamond from $s$, because different orderings of $!m$ and $!o$ will result in different channel orderings.
Therefore $\Conf^a_b(s')$ cannot hold, which makes the implication hold trivially.
\qed
\end{proof}

\noindent\textbf{Theorem \ref{the:confmerge}.}
\textit{
Let $F = \langle P, C, M \rangle$ and $F' = \langle P, C', M \rangle$ be FIFO systems such that $F \cgtc F'$.
Let $B_F = \langle S, s_0, L, \trans{} \rangle$ and $B_{F'} = \langle S', s_0', L, \ptrans{} \rangle$.
Let $a, b \in L$.
If not $\AC{a}{b}$ and not $a \equiv_? b$, then confluence of $a$ and $b$ is merge-preserved.
}
\begin{proof}
This follows from Lemma \ref{lem:statecorr}, \ref{lem:reachsplit}, \ref{lem:conftau}, \ref{lem:confoc}, \ref{lem:confsc} and \ref{lem:confccmerge}.
\qed
\end{proof}

\noindent\textbf{Theorem \ref{the:confsplit}.}
\textit{
Let $F = \langle P, C, M \rangle$ and $F' = \langle P, C', M \rangle$ be FIFO systems such that $F \cgtc F'$.
Let $B_F = \langle S, s_0, L, \trans{} \rangle$ and $B_{F'} = \langle S', s_0', L, \ptrans{} \rangle$.
Let $a, b \in L$.
If $\SC{a}$, $\SC{b}$ and $S \unre S'$, then confluence of $a$ and $b$ is split-preserved.
}
\begin{proof}
This follows from Lemma \ref{lem:conftau}, \ref{lem:confoc}, \ref{lem:confsc} and \ref{lem:confccsplit}.
\qed
\end{proof}

\end{document}